\documentclass[useAMS,usenatbib]{mn2e}

\usepackage{amsmath}
\usepackage{amssymb}
\usepackage{graphicx}
\usepackage{graphics}
\usepackage{float}
\usepackage{epstopdf}
\usepackage{multicol}
\usepackage{breakurl}
\usepackage{textcomp}
\usepackage{enumitem}
\usepackage{natbib}
\usepackage{subfigure}
\usepackage[bf,labelsep=period,font=small]{caption}
\usepackage{subfloat}
\usepackage{subfigure}
\usepackage{gensymb}
\usepackage{supertabular}
\usepackage{longtable,ltxtable}
\usepackage{afterpage}
\usepackage[flushleft]{threeparttable}
\title[MOBSTER II. Rotationally variable A stars]{MOBSTER - II. Identification of rotationally variable A stars observed with \emph{TESS} in Sectors 1 to 4}
\author[J. Sikora et al.]
	{J.~Sikora,$^{1,2}$ A.~David-Uraz,$^3$ S.~Chowdhury,$^4$ D.~M.~Bowman,$^5$ G.~A.~Wade,$^2$\newauthor V.~Khalack,$^6$ O.~Kobzar,$^6$ O.~Kochukhov,$^7$ C.~Neiner,$^8$ E.~Paunzen,$^9$\\
$^{1}$Department of Physics, Engineering Physics \& Astronomy, Queen's University, Kingston, ON K7L 3N6, Canada\\
$^{2}$Department of Physics and Space Science, Royal Military College of Canada, PO Box 17000 Kingston, ON K7K 7B4, Canada\\
$^{3}$Department of Physics and Astronomy, University of Delaware, Newark, DE 19716, USA\\
$^{4}$Nicolaus Copernicus Astronomical Center, Bartycka 18, PL-00-716 Warsaw, Poland\\
$^{5}$Instituut voor Sterrenkunde, KU Leuven, Celestijnenlaan 200D, 3001 Leuven, Belgium\\
$^{6}$D{\'e}partement de Physique et d'Astronomie, Universit{\'e} de Moncton, Moncton, NB E1A 3E9, Canada\\
$^{7}$Department of Physics and Astronomy, Uppsala University, SE-751 20 Uppsala, Sweden\\
$^{8}$LESIA, Paris Observatory, PSL University, CNRS, Sorbonne Universit\'e, Universit\'e de Paris, 5 place Jules Janssen, \\\hspace{1cm}92195 Meudon, France\\
$^{9}$Department of Theoretical Physics and Astrophysics, Masaryk University, Kotl{\'a}\u rsk{\'a} 2, 611 37 Brno, Czech Republic}

\begin{document}

\date{Accepted 2019 May 21}

\pagerange{\pageref{firstpage}--\pageref{lastpage}} \pubyear{2019}

\maketitle

\label{firstpage}

\begin{abstract}
Recently, high-precision optical 2~min cadence light curves obtained with \emph{TESS} for targets 
located in the mission's defined first four sectors have been released. The majority of these 
high-cadence and high-precision measurements currently span $\sim28\,{\rm d}$, thereby 
allowing periodic variability occurring on timescales $\lesssim14\,{\rm d}$ to potentially be detected. 
Magnetic chemically peculiar (mCP) A-type stars are well known to exhibit rotationally modulated 
photometric variability that is produced by inhomogeneous chemical abundance distributions in their 
atmospheres. While mCP stars typically exhibit rotation periods that are significantly longer than those 
of non-mCP stars, both populations exhibit typical periods $\lesssim10\,{\rm d}$; therefore, the early 
\emph{TESS} releases are suitable for searching for rotational modulation of the light curves of both 
mCP and non-mCP stars. We present the results of our search for A-type stars that exhibit variability in 
their \emph{TESS} light curves that is consistent with rotational modulation based on the first two data 
releases obtained from sectors 1 to 4. Our search yielded $134$ high-probability candidate rotational 
variables -- $60$ of which have not been previously reported. Approximately half of these stars are 
identified in the literature as Ap (mCP) stars. Comparisons between the subsample of high-probability 
candidate rotationally variable Ap stars and the subsample of stars that are not identified as Ap reveal 
that the latter subsample exhibits statistically (i) shorter rotation periods and (ii) significantly 
lower photometric amplitudes.
\end{abstract}


\section{Introduction}\label{sect:intro}

One fundamental property that distinguishes main sequence (MS) A-type stars with masses 
$1.5<M<3\,M_\odot$ from their lower-mass ($M\lesssim1.5\,M_\odot$) counterparts is the incidence rate of 
inhomogeneous surface brightness distributions, which are broadly referred to as star spots. Contrary to 
MS A-type stars, essentially all late-F, G-, and K-type MS stars exhibit bright and/or dim star spots 
\citep{Cameron2002,Berdyugina2005,Ermolli2007}. The relatively small fraction of MS A-type 
stars that do host qualitatively similar surface structures are known as $\alpha^2$~CVn variables and 
are identified based on detections of low-frequency photometric and spectroscopic variability 
\citep{GCVS2017}. While the origins of these surface structures differ, in both the spotted low-mass MS 
star and $\alpha^2$~CVn cases, the observed photometric variability is understood to be a consequence of 
the star's rotation: the star's brightness is modulated as surface features appear and disappear from 
view.

It is well established that $\alpha^2$~CVn variable stars are magnetic chemically peculiar (mCP) Ap 
stars \citep[e.g.][]{Babcock1952}. Furthermore, the origin of surface brightness inhomogeneities that 
lead to photometric rotational modulation in $\alpha^2$~CVn stars can be traced to the presence of 
strong, organized magnetic fields that are visible at the stellar surface 
\citep{Krticka2007,Krticka2009,Krticka2015}. The mCP stars account for approximately $10$~per~cent of all 
MS A-type stars with masses $\sim3\,M_\odot$ \citep{Wolff1968,Auriere2007,Sikora2019}. They host strong, 
stable, and organized magnetic fields with strengths $\sim0.1-30\,{\rm kG}$ 
\citep{Babcock1960,Landstreet1982}. Recently, a small number of non-mCP MS A-type stars have been found 
to host ultra-weak (dubbed ``Vega-like" after the prototype of the population) fields with strengths 
$\lesssim1\,{\rm G}$ \citep{Lignieres2009,Petit2011,Blazere2016a}. It is has been suggested that these 
stars may represent a new class of magnetic MS A-type stars that are distinct from the mCP stars 
\citep{Lignieres2014}.

The \emph{Kepler} \citep{Borucki2010}, K2 \citep{Howell2014}, and CoRoT \citep{Auvergne2009} space-based 
photometry missions have resulted in the identification of a large number of MS A-type stars exhibiting 
variability that is consistent with rotational modulation \citep{Balona2011,Paunzen2015,Bowman2018a}. In 
particular, \citet{Balona2013} reported the discovery that $\sim40$~per~cent of MS A-type stars observed 
with \emph{Kepler} may exhibit photometric rotational variability. In addition to the rotational 
variables, \citet{Balona2012,Balona2013} also identified (i) a number of apparently flaring A-type stars 
and (ii) unexplained periodogram features that are characterized by nearly coincident low-frequency 
diffuse and narrow peaks \citep[e.g. Fig. 6 of][]{Balona2013}.

\citet{Pedersen2017} carried out a detailed analysis of 33 of the A-type stars that reportedly 
exhibit evidence of flares in their \emph{Kepler} light curves. They confirmed the presence of flares in 
27 of these stars' light curves but found that the flares associated with 14 of these cases may be 
attributed to contamination from nearby sources. Ultimately, these authors concluded that the flares 
identified by \citet{Balona2012,Balona2013} are likely not intrinsic to the A stars themselves. In the 
case of the unexplained periodogram features, \citet{Balona2014} suggested that the diffuse peaks may be 
caused by spots appearing on the surface of a differentially rotating A star while the narrow peaks are 
caused by close-in Jupiter-sized planets. However, \citet{Saio2018} argue for a more physically-justified 
explanation in that the features in the amplitude spectra of many A stars are evidence of Rossby waves 
\citep[also known as r modes,][]{Papaloizou1978}, which can occur in rotating $\gamma$ Dor stars. 
Observational evidence of this phenomenon has been previously identified by \citet{VanReeth2016}.

If the variability identified by \citet{Balona2013} is indeed produced by spots on the surfaces of 
rotating A-type stars, then what is the origin of the spots? Such variability in these particular 
objects is normally attributed to the presence of magnetic fields that are visible at the surface. The 
reported $\sim40$~per~cent incidence rate of rotationally variable A-type stars is several times higher 
than the incidence rate of strongly magnetic (mCP) stars; therefore, it is unlikely that all of these stars 
are strongly magnetic. Considering that surface spots have been detected on Vega \citep[which hosts an 
ultra-weak field,][]{Bohm2015} and that such spots may lead to photometric variability, it is plausible that a 
large number of the A-type stars that were found to exhibit rotational modulation by \citet{Balona2013} 
may host ultra-weak fields. Recently, \citet{Cantiello2019} explored the observational consequences of 
convection that might produce dynamo magnetic fields at or near the surfaces of fast-rotating A- and 
B-type stars. Referring to similar discussions presented by \citet{Cantiello2011} in the context of more 
massive stars, the authors note that such magnetic fields are expected to result in both ultra-weak fields 
and coincident spots characterized by low temperature contrasts ($\sim10\,{\rm K}$) being present at the 
stellar surface. They predict that, while all A- and late B-type stars should host these features, they are 
likely largely undetectable. The fraction of stars for which the temperature spots can be expected to be 
detected depends on the size and number of spots \citep{Kochukhov2013}, which is not well constrained.

The MS A-type stars as defined in this study partly overlap with the mercury-manganese (HgMn) group of 
CP stars, which exhibit a distinct and unique rotational variability. HgMn stars have masses between 
$2.5\,M_\odot$ and $\sim5\,M_\odot$ and effective temperatures as low as $10\,000\,{\rm K}$ 
\citep{Ghazaryan2016}. These stars show no evidence of either global or strong complex magnetic fields 
\citep{Auriere2010,Makaganiuk2010,Kochukhov2013a}, with the best upper limits on the longitudinal field 
component currently in the $1-10\,{\rm G}$ range. Nevertheless, many of these stars show rotational line 
profile variability indicative of low-contrast, large-scale inhomogeneous surface abundance distributions 
\citep{Adelman2002,Kochukhov2005,Folsom2010,Briquet2010,Makaganiuk2012}. Unlikely stationary spots on 
magnetic Ap stars, surface structures on HgMn stars evolve on time scales from months to years 
\citep{Kochukhov2007,Korhonen2013}, suggesting a different underlying physical mechanism of their 
formation. A low-amplitude photometric modulation associated with chemical spots on HgMn stars is 
undetectable from the ground but can be identified using high-precision space photometric data 
\citep{Alecian2009,Balona2011a,Hodgson2017}.

The MOBSTER collaboration \citep[Magnetic $\rm{OB[A]}$ Stars with \emph{TESS}: probing their Evolutionary and 
Rotational properties;][]{DavidUraz2019} is focused on advancing our understanding 
of stellar magnetism of intermediate- and high-mass stars using \emph{TESS} observations along with 
spectroscopic and spectropolarimetric follow-up observations. The primary goal of the study presented 
here, which is the second of a series of publications by the MOBSTER collaboration, is to 
identify new candidate rotational variable MS A-type stars based on the release of \emph{TESS} 
observations. These stars are hypothesized to host either strong magnetic fields (similar to those 
associated with mCP stars) or ultra-weak fields \citep[similar to that found on Vega,][]{Lignieres2009}. 
Since these stars are generally much brighter than those detected by \emph{Kepler}, these identifications 
will serve as the basis for spectropolarimetric surveys designed to detect such magnetic fields, as was 
proven a successful strategy with stars observed by K2 \citep[e.g.][]{Buysschaert2018}.

In Sect. \ref{sect:obs} we describe the \emph{TESS} observations on which our study is based. In Sect. 
\ref{sect:sample} we describe our sample of A-type stars and the methods by which it was constructed. 
In Sect. \ref{sect:rot_mod} we discuss how our search for rotationally modulated \emph{TESS} light 
curves was carried out and present the results of this search. In Sect. \ref{sect:fund_param}, we 
present the fundamental parameters (i.e. effective temperatures, radii, masses, etc.) associated with 
our sample and describe the way in which they were derived. In Sect. \ref{sect:targets}, we discuss some 
of the more noteworthy targets identified as candidate rotational variables. Finally, in Sect. 
\ref{sect:disc} we summarize and discuss the results of this study.

\section{Observations}\label{sect:obs}

\emph{TESS} is optimized to detect planetary transit signatures in light curves of MS dwarf stars having 
$I_C$ magnitudes of approximately $4-13$ \citep{Ricker2015}. For typical MS A-type stars, which are the 
focus of this study, the $I_C$ limits correspond to Johnson $V$ magnitudes of approximately $3-12$. The 
passband of the filter used by the onboard photometer has an effective wavelength of 
$\approx7\,500\,{\rm \AA}$ and a width of $\approx4\,000\,{\rm \AA}$ \citep{Sullivan2015}. In this study 
we used the $2\,{\rm min}$ PDC\_SAP light curves processed by the \emph{TESS} Science Team. These light 
curves are available from the Mikulski Archive for Space Telescopes 
(MAST)\footnote{https://archive.stsci.edu/tess/}. We refer the reader to \citet{Jenkins2016} for a 
description of the pipeline that produces these light curves.

The \emph{TESS} data considered in this study consist of targets located in sectors 1 to 4. These 
sectors are in the southern ecliptic and contain targets with right ascension (RA) values of 
${\rm RA}<131$~degrees and ${\rm RA}>308$~degrees and declination (Dec) values of 
$-85<{\rm Dec}<+12$~degrees. The observations have been obtained over a period of $\approx4\,{\rm months}$ 
from July 25 to November 14, 2018. Sectors 1 and 2, sectors 2 and 3, and sectors 3 and 4 partially 
overlap where ${\rm Dec}\lesssim-30$~degrees; the light curves associated with the targets found in only 
one sector (the majority of the targets) span $28\,{\rm d}$ while those associated with targets found 
in two sectors span $56\,{\rm d}$. The brightness measurements exhibit typical uncertainties of 
$\lesssim800\,{\rm parts~per~million}$ (ppm).

\section{Sample}\label{sect:sample}

\emph{TESS} light curves of $44\,371$ targets in sectors 1 to 4 were made available during the first two 
data releases. In order to identify the A-type stars within this sample, we first cross-referenced this 
list with the SIMBAD astronomical database\footnote{http://simbad.u-strasbg.fr/simbad/} for available 
spectral types associated with those targets. No spectral types could be found for $20\,098$ of the 
$44\,371$ observed targets while a further $5\,083$ targets could not be found within SIMBAD. We 
identified $1\,715$ A-type stars from the $19\,190$ \emph{TESS} targets with available spectral types.

We attempted to identify the A-type stars in the subset of $25\,181$ stars ($20\,098+5\,083$) without 
available spectral types using their effective temperatures ($T_{\rm eff}$). The \emph{TESS} Input 
Catalogue \citep[TIC,][]{Stassun2018} includes $T_{\rm eff}$ values that are either derived from 
$(V-K_{\rm S})-T_{\rm eff}$ calibrations \citep{Casagrande2008,Huang2015} or are taken from published 
spectroscopic surveys. Typical A-type stars exhibit $7\,000\lesssim T_{\rm eff}\lesssim10\,000\,{\rm K}$; 
however, the colour-$T_{\rm eff}$ calibrations used for the majority of the $T_{\rm eff}$ values 
reported in the TIC are only applicable to stars having $T_{\rm eff}\leq9\,755\,{\rm K}$. Out of the 
subset of $25\,181$ observed stars without available spectral types that are listed in the TIC, $228$ 
exhibit $7\,000\leq T_{\rm eff}\leq9\,755\,{\rm K}$ and another $228$ have not been assigned 
$T_{\rm eff}$ values (i.e. their $T_{\rm eff}$ values could not be estimated using the 
$(V-K_{\rm S})-T_{\rm eff}$ calibrations and no spectroscopically determined $T_{\rm eff}$ values are 
available).

Johnson $B$ and $V$ magnitudes are available for $93$ of the $228$ stars without $T_{\rm eff}$ values 
listed in the TIC. We estimated their $T_{\rm eff}$ values using the $(B-V)-T_{\rm eff}$ calibration 
published by \citet{Gray2005_SP} for MS stars. This colour-$T_{\rm eff}$ calibration is reasonably 
sensitive to stars with $T_{\rm eff}\lesssim10\,000\,{\rm K}$; the scatter associated with the 
calibration is estimated to be $\approx2-5$~per~cent. The $(B-V)-T_{\rm eff}$ calibration yielded $19$ 
stars having $7\,000\leq T_{\rm eff}\leq10\,000\,{\rm K}$ and are therefore considered to be candidate 
A-type stars.

In summary, out of the $44\,371$ targets included in the first two \emph{TESS} data releases, we identified 
a total of $1\,962$ A-type stars based on (i) published spectral types ($1\,715$ stars), (ii) the 
$(V-K_{\rm S})-T_{\rm eff}$ calibrations used in the TIC ($228$ stars), and (iii) $(B-V)-T_{\rm eff}$ 
calibrations ($19$ stars). We note that by adopting a minimum $T_{\rm eff}$ value of $7\,000\,{\rm K}$ 
in our search for A-type stars within the subsample of those stars without available spectral types, a 
number of cooler A-type stars will likely have been missed. While this is unlikely to significantly 
impact the broader findings of our study, it may have an impact on the results associated with the coolest 
and lowest mass A-type stars in our sample.

Johnson $V$ magnitudes for all but one of the $1\,962$ A-type stars included in our sample are listed 
in the TIC. The sample exhibits a median $V$ magnitude of $8.7\,{\rm mag}$ and ranges from $16.4$ to 
$2.9\,{\rm mag}$. Distances ($d$) derived from the parallax angles associated with the second Gaia Data 
Release (DR2) \citep{Gaia2016,Holl2018,Bailer-Jones2018} for $1\,906$ of the $1\,962$ sample stars are 
available; based on these values, the majority of the stars in the sample are relatively nearby with 
$90$~per~cent having $d\lesssim440\,{\rm pc}$. Out of the $1\,962$ stars in our sample, $363$ are 
located in overlapping sectors and thus, their light curves have a temporal baseline of $56\,{\rm d}$ 
while the remaining $1\,599$ appear only in one sector and have temporal baselines of $28\,{\rm d}$.

\section{Rotational Modulation}\label{sect:rot_mod}

\subsection{Search criteria}\label{sect:search_criteria}

We carried out a search for rotationally modulated variability in the set of \emph{TESS} light curves 
by adopting the main criteria outlined by \citet{Balona2011,Balona2013}: (i) the frequency spectrum of 
the light curve (i.e. the periodogram) must exhibit a low frequency peak ($f_1\lesssim3\,{\rm d^{-1}}$) 
that plausibly corresponds to the star's rotational frequency and (ii) this peak must be accompanied by 
a first harmonic ($f_2=2f_1$). The first criterion is based on the fact that the critical rotation 
frequency of a zero age MS (ZAMS) A-type star ranges from $f_{\rm crit}\approx3-3.7\,{\rm d^{-1}}$. As 
described by \citet{Balona2013}, the second criterion is adopted as a means of minimizing the number of 
stars exhibiting low frequency signals that may be caused by pulsations rather than rotational 
modulation \citep[e.g. $\gamma$~Dor stars and slowly pulsating B-type 
stars,][]{Cousins1989,Waelkens1991,Cat2002,Henry2003}.

As a result of the second criterion, rotational variables may not be detected in our study if the 
amplitude of the first harmonic ($A_2$) is low. The ratio of the amplitude of the rotational 
frequency ($A_1$) to $A_2$ depends largely on the distribution of co-rotating surface structures (i.e. 
spots) and on the inclination angle, $i$, of the star's rotational axis. We attempted to evaluate the 
selection effect that the second criterion has on our sample of identified rotational variables by using 
the analytical spot model developed by \citet{Eker1994}. The model consists of a rigidly rotating star 
having one or more circular surface spots positioned at various longitudes and latitudes. The inclination 
of the star's rotation axis, the manner in which limb darkening is treated (i.e. whether it is 
neglected or described by a linear or quadratic limb darkening law), and the spot radii and contrast 
values are input parameters that can be modified in this model.

The expected range of the ratio $A_1/A_2$ was estimated by carrying out several Monte Carlo (MC) 
simulations. We generated $500$ models each for cases in which between 1 and 5 spots are present on 
the star's surface. For each model, the star's $i$ value, spot longitudes, latitudes, and angular radii 
(ranging from $2$ to $30$~degrees) were randomly assigned; for those models with more than one 
spot, the same radius was adopted for each spot. We used a linear limb darkening law with a coefficient 
of $u=0.589$, which is computed by \citet{Diaz-Cordoves1995} in the $V$ band for a $7\,000\,{\rm K}$ star 
with $\log{g}=4.0\,{\rm (cgs)}$. We adopted a rotation period of $2.5\,{\rm d}$ and the light curves 
were then calculated during $40\,{\rm d}$ ($16$ rotational cycles) with a cadence of $2\,{\rm min}$. 
Lomb-Scargle (LS) periodograms \citep{Lomb1976,Scargle1982,Press2007} were calculated for each synthetic 
light curve (the same technique was employed in the analysis of the \emph{TESS} light curves, as 
described below) from which the values of $A_1$ and $A_2$ were extracted in units of magnitude.

\begin{figure}
	\centering
	\includegraphics[width=0.8\columnwidth]{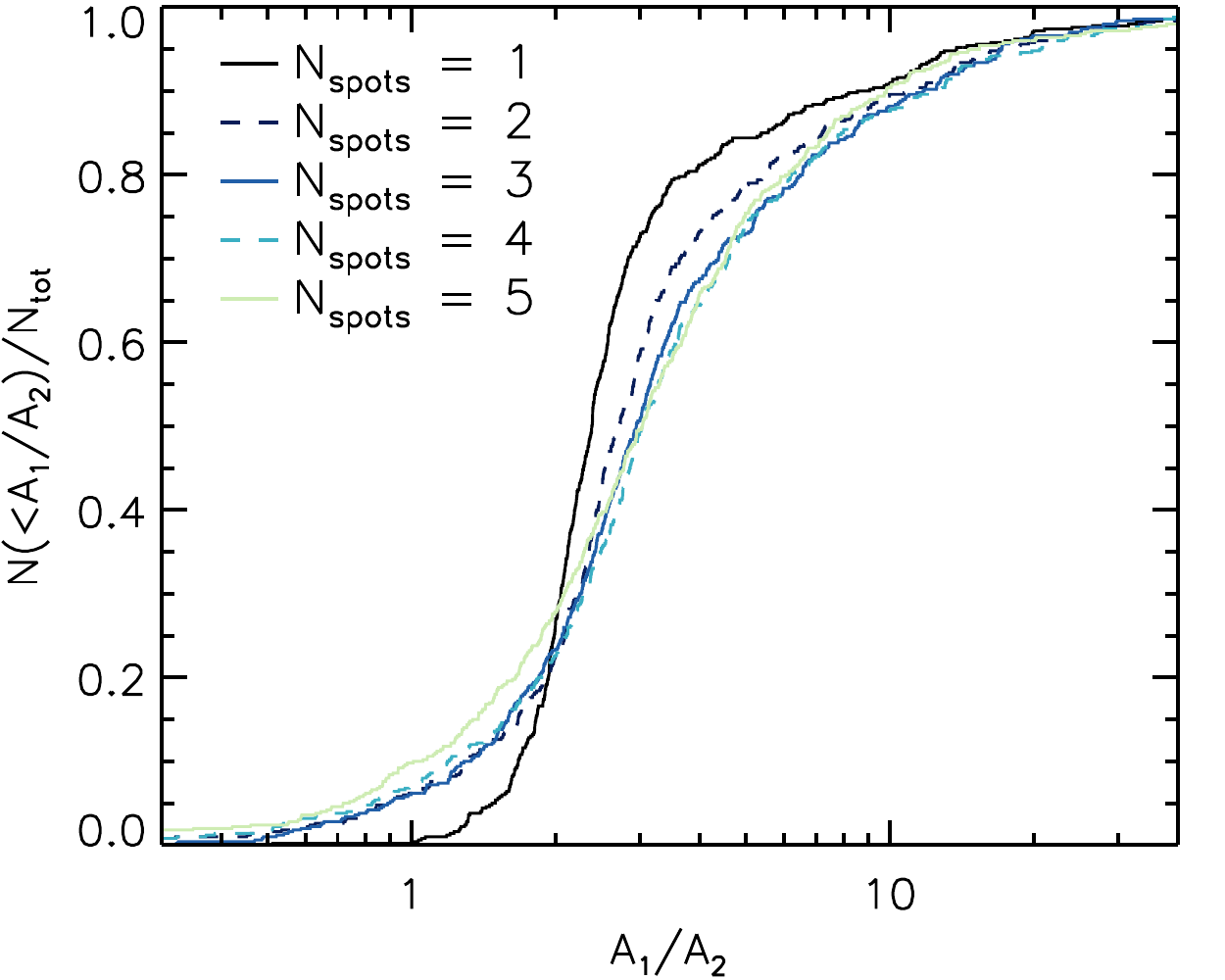}
	\caption{Cumulative distribution functions corresponding to $A_1/A_2$ based on the spot model 
	published by \citet{Eker1994}. The distributions for models with 1, 2, 3, 4, and 5 spots are shown 
	increasing in brightness from dark blue to light blue and alternating between solid and dashed lines.}
	\label{fig:mc_spot_a1_a2}
\end{figure}

In Fig. \ref{fig:mc_spot_a1_a2} we show the cumulative distribution functions for $A_1/A_2$ associated 
with the MC simulations carried out with 1 to 5 spots. The results suggest that stars having 
multiple spots are more likely to exhibit $A_1/A_2$ ratios that are $\gtrsim2$ compared to stars with 
single spots. This implies that stars with complex spot distributions are less likely to satisfy the 
second criterion in our search for rotational modulation since, depending on the value of $A_1$, the 
value of $A_2$ is more likely to fall below the detection threshold. It is also noted that a 
non-negligible fraction ($\sim10\,$per~cent) of the models exhibiting more than one spot have 
$A_1/A_2<1$ and thus, the rotation frequency does not necessarily correspond to the periodogram peak 
with the largest amplitude.

The two criteria used to identify rotationally modulated light curves are also frequently satisfied 
by eclipsing binaries (EBs) and ellipsoidal variables (EVs). EBs are easily identified based on the 
presence of primary and secondary eclipses in their light curves; however, EVs can exhibit highly 
sinusoidal light curves similar to those produced by rotational modulation 
\citep[e.g. Fig. 10 of][]{Smalley2014}. In order to reduce the number of EVs misclassified as rotational 
variables, we rejected any non-Ap rotational variable candidates with light curves exhibiting a deep 
local minimum followed by a shallower local minimum that are both bracketed by the system's maximum 
brightness. These characteristic features of EVs are often easy to identify when the light curve is 
phased by the system's orbital period since the two minima have a phase separation of $0.5$. 
Furthermore, the distances separating the two components of an EV system are relatively low and the 
inclination of the orbital plane is relatively high; therefore, radial velocity variations are often 
large and easy to detect \citep[e.g.][]{Matson2017}. We searched the literature for any reported 
detections of radial velocity variations from which orbital periods have been derived. Those candidate 
rotational variables with reported orbital periods consistent with the photometric periods that were 
originally attributed to the star's rotation were rejected.

The last step of the search process involved classifying the identified candidate rotational 
variables as either low- or high-probability candidates. In a number of cases, the rotation and first 
harmonic peaks were accompanied by additional low-frequency peaks with comparable amplitudes \citep[e.g. 
$\gamma$ Dor stars,][]{Henry2003}. These stars were considered to be low-probability candidate 
rotational variables due to our inability to distinguish between those peaks that are caused by 
rotational modulation and those which are caused by pulsations; all other candidates were assigned a 
high-probability status.

\begin{figure*}
	\centering
	\includegraphics[width=2.0\columnwidth]{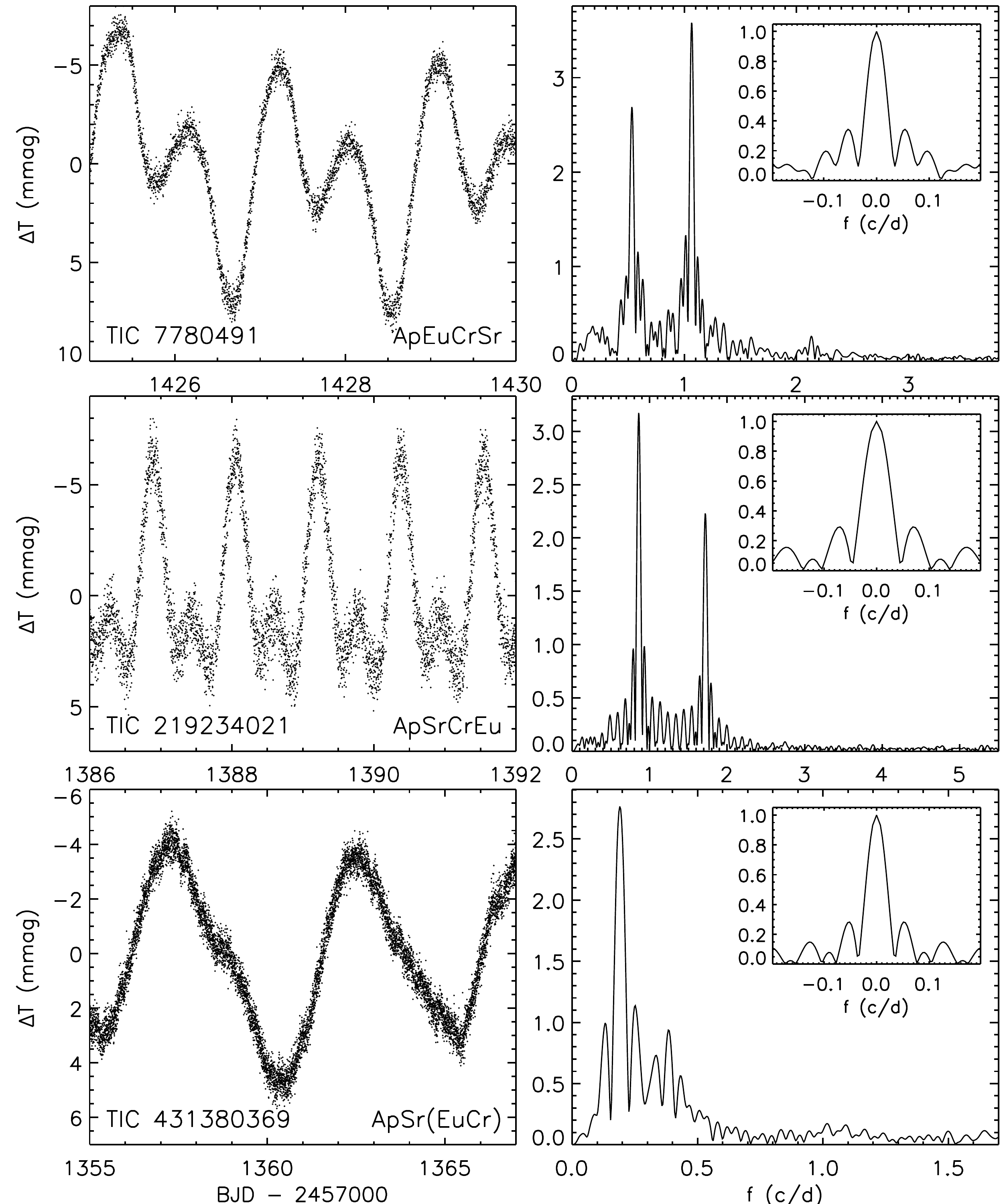}
	\caption{Examples of \emph{TESS} light curves associated with Ap stars that are found to exhibit 
	variability that is consistent with a rotational origin. \emph{Left:} Subsamples of the full light 
	curves. \emph{Right:} The Lomb-Scargle periodograms derived from the light curves. The rotational 
	frequencies believed to be the stars' rotational frequencies are apparent along with the first 
	harmonic. The spectral window function is shown in the inset at upper right.}
	\label{fig:ex_LC_LS_Ap}
\end{figure*}

\begin{figure*}
	\centering
	\includegraphics[width=2.0\columnwidth]{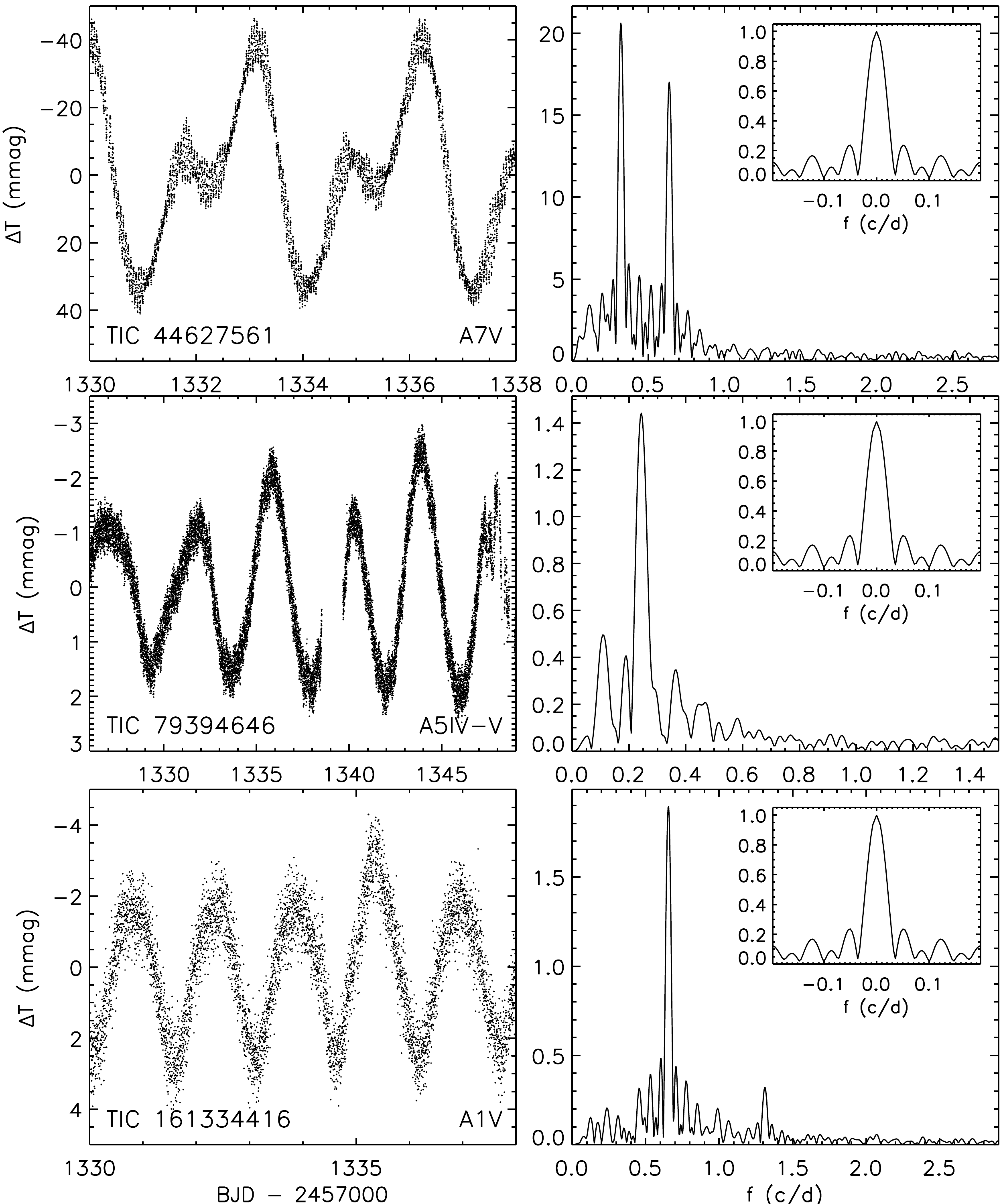}
	\caption{Same as Fig. \ref{fig:ex_LC_LS_Ap} but for non-Ap stars. Note that TIC~44627561 (top) 
	exhibits high-frequency $\sim20\,{\rm d}^{-1}$ variability that is typical of $\delta$~Scuti stars.}
	\label{fig:ex_LC_LS_A}
\end{figure*}

\subsection{Application of the search criteria to the \emph{TESS} sample}

We found that the signatures of rotational modulation in the light curves that were reduced and made 
publicly available by the \emph{TESS} team could be more clearly identified after they were 
post-processed.  This involved manually removing obvious outliers and detrending instrumental 
systematics by subtracting a low-order polynomial. We note that this detrending process may remove 
long-term trends described by periods that are approximately one or more times as long as the total 
light curve's timespan. First- or second-order polynomial fits are unlikely to remove short period signals 
that are strictly periodic; however, signals that exhibit variations between consecutive cycles (e.g. 
such as those which might be associated with evolving star spots) are more susceptible to being 
removed.

We calculated LS periodograms of these post-processed light curves using an oversampling factor of ten 
to ensure that peaks in a periodogram were well resolved. Statistically significant peaks were then 
identified using an iterative pre-whitening procedure similar to that described by \citet{Morel2011}. 
For each iteration, the frequency of the maximum-amplitude peak ($f_{\rm max}$) appearing in the 
periodogram was identified and a first-order sinusoidal function having frequency $f_{\rm max}$ (i.e. 
$A\sin{[2\pi f_{\rm max}t+\phi]}$ where $A$, $t$, and $\phi$ correspond to the semi-amplitude, time, and 
phase) was fit to the light curve with $A$ and $\phi$ as free parameters. The fit was then subtracted 
from the light curve and the LS periodogram was recalculated. This procedure was repeated until either 
$50$ frequencies had been extracted or until no peaks having a statistical significance $\geq5\,\sigma$ 
were found \citep[the significance is evaluated according to Eqn. 13.8.7 of][]{Press2007}. Uncertainties 
in all the extracted frequencies and their associated amplitudes were estimated using the prescriptions 
described by \citet{Montgomery1999}, which involve the root mean square deviation (RMSD) of the noise 
inherent in the light curve ($\sigma[m]$). We estimated $\sigma(m)$ by taking the RMSD of the light 
curve once all of the statistically significant signals had been removed.

After extracting the frequencies as described above, we carried out an automated routine designed to 
identify rotation ($f_1$) and first harmonic ($f_2$) frequency pairs. The first harmonic is given by 
$f_2=2f_1\pm\Delta f$ where $\Delta f$ is defined as the half-width at half-maximum of the central peak of 
each light curve's spectral window function corresponding to a typical value of $0.02\,{\rm d^{-1}}$. The 
$1\,695$ light curves exhibiting low frequency peaks with accompanying first harmonics were then inspected 
visually; those targets with obvious EB, EV, or pulsational (e.g. RR~Lyrae stars) signatures were removed 
from the list of candidate rotational variables. The remaining candidates were then assigned either a 
low-probability or high-probability candidate rotational variable classification based on the appearance 
of additional low-frequency, high-amplitude peaks (described in Sect. \ref{sect:search_criteria}).

Having identified periodogram peaks that are likely associated with each star's rotation period 
($P_{\rm rot}=1/f_1$), we refined the final $P_{\rm rot}$ values using a Levenberg-Marquardt 
least-squares algorithm. This involved fitting the light curves to a sinusoidal function that includes the 
first four harmonics:
\begin{equation}\label{eqn:multi_order_fit}
	\Delta T(t)=a_0+\sum_{n=1}^{n=5}a_n\sin{(2\pi n t/P+\phi_n)}
\end{equation}
where $a_0$, $a_n$, $\phi_n$, and $P$ are free parameters. The $a_1$ and $P$ parameters are 
assigned initial guesses of half the range of $\Delta T(t)$ and $1/f_1$, respectively, while all other 
parameters are assigned initial guesses of zero. The uncertainties in each of the fitting parameters were 
estimated using a residual bootstrapping method. The refined rotation periods associated with the 
high-probability candidates along with the maximum amplitudes ($\Delta T_{\rm max}$, i.e. the largest 
$a_n$, which always corresponds to either $a_1$ or $a_2$) associated with the sinusoidal fitting function 
are listed in Table \ref{tbl:vmag_Prot}.

In Fig. \ref{fig:ex_LC_LS_Ap}, we show sample light curves and their associated periodograms of 
several Ap stars identified as high-probability candidate rotational variables. Similarly, Fig. 
\ref{fig:ex_LC_LS_A} shows examples of stars that are classified as non-chemically peculiar (i.e. their 
spectral types do not contain a `p' classification) in the literature and are also considered to be 
high-probability candidate rotational variables. The search yielded $134$ high-probability and $126$ 
low-probability candidate rotational variables, respectively. This corresponds to an incidence rate of 
$6.8\pm0.6$~per~cent. Including those stars identified as low-probability rotational variables 
increases this incidence rate to $13.3\pm0.9$~per~cent. Out of the $134$ high-probability candidate 
rotational variables, $76$ are classified as Ap stars, $3$ are classified as Am stars, $7$ have luminosity 
classes of II or III (i.e. they have likely evolved off the MS), and $48$ are either classified as 
non-chemically peculiar MS stars or are unclassified. 

\begin{figure}
	\centering
	\includegraphics[width=1.0\columnwidth]{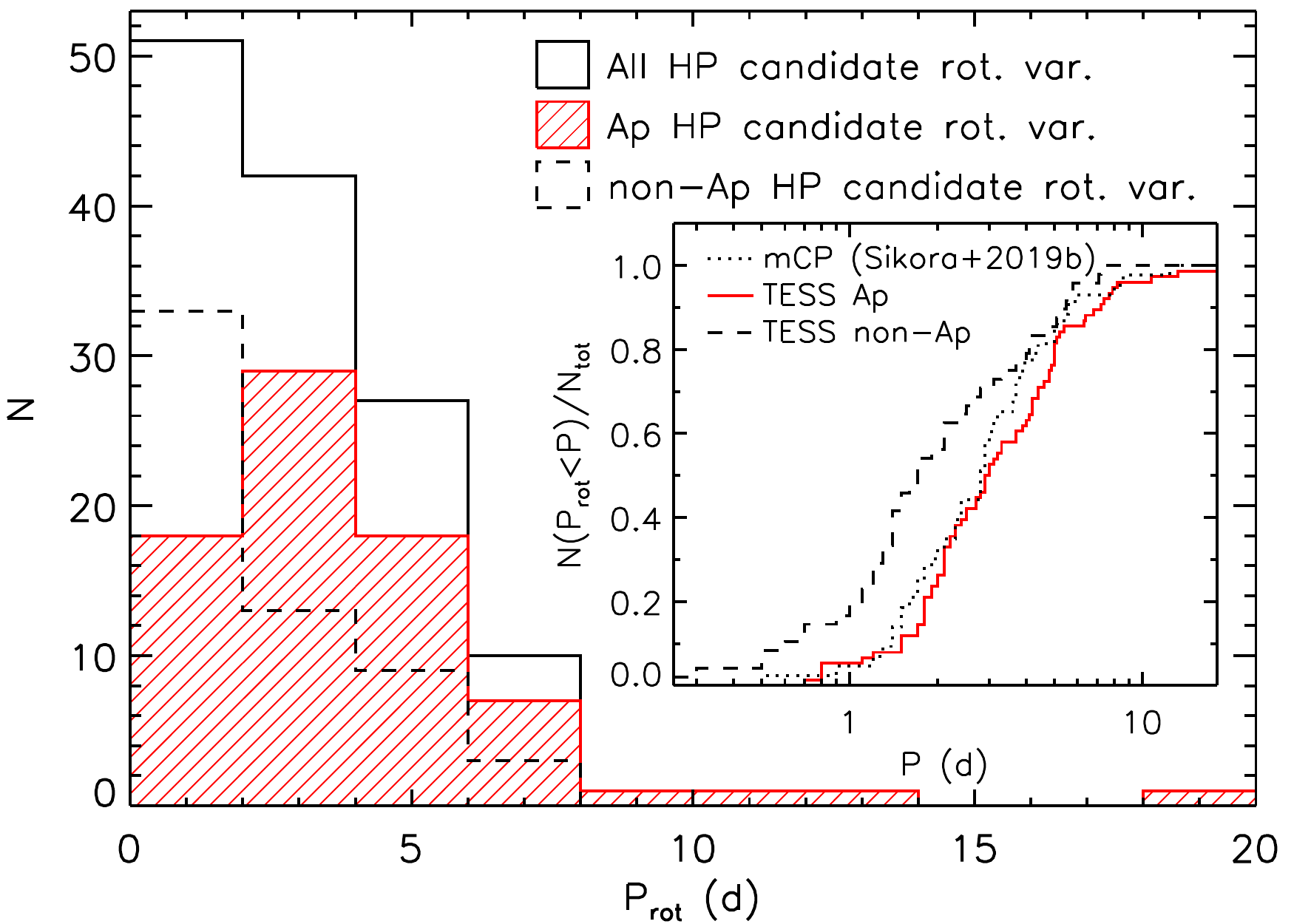}
	\caption{Distributions of the rotation periods ($P_{\rm rot}$) obtained from the \emph{TESS} light 
	curves associated with the high-probability (HP) rotational variable candidates for the 
	total sample (solid black), subsample of Ap stars (red hatched), and subsample of non-Ap stars 
	(dashed black). \emph{Inset:} The cumulative distribution functions associated with the $P_{\rm rot}$ 
	values of the \emph{TESS} stars not classified as being Ap (dashed black), the \emph{TESS} stars 
	classified as Ap (solid red), and the confirmed mCP stars having $P_{\rm rot}<15\,{\rm d}$ 
	($90$~per~cent of the sample of mCP stars) included in the survey carried out by \citet{Sikora2019a} 
	(dotted black).}
	\label{fig:Prot_hist}
\end{figure}

\begin{figure}
	\centering
	\includegraphics[width=1.0\columnwidth]{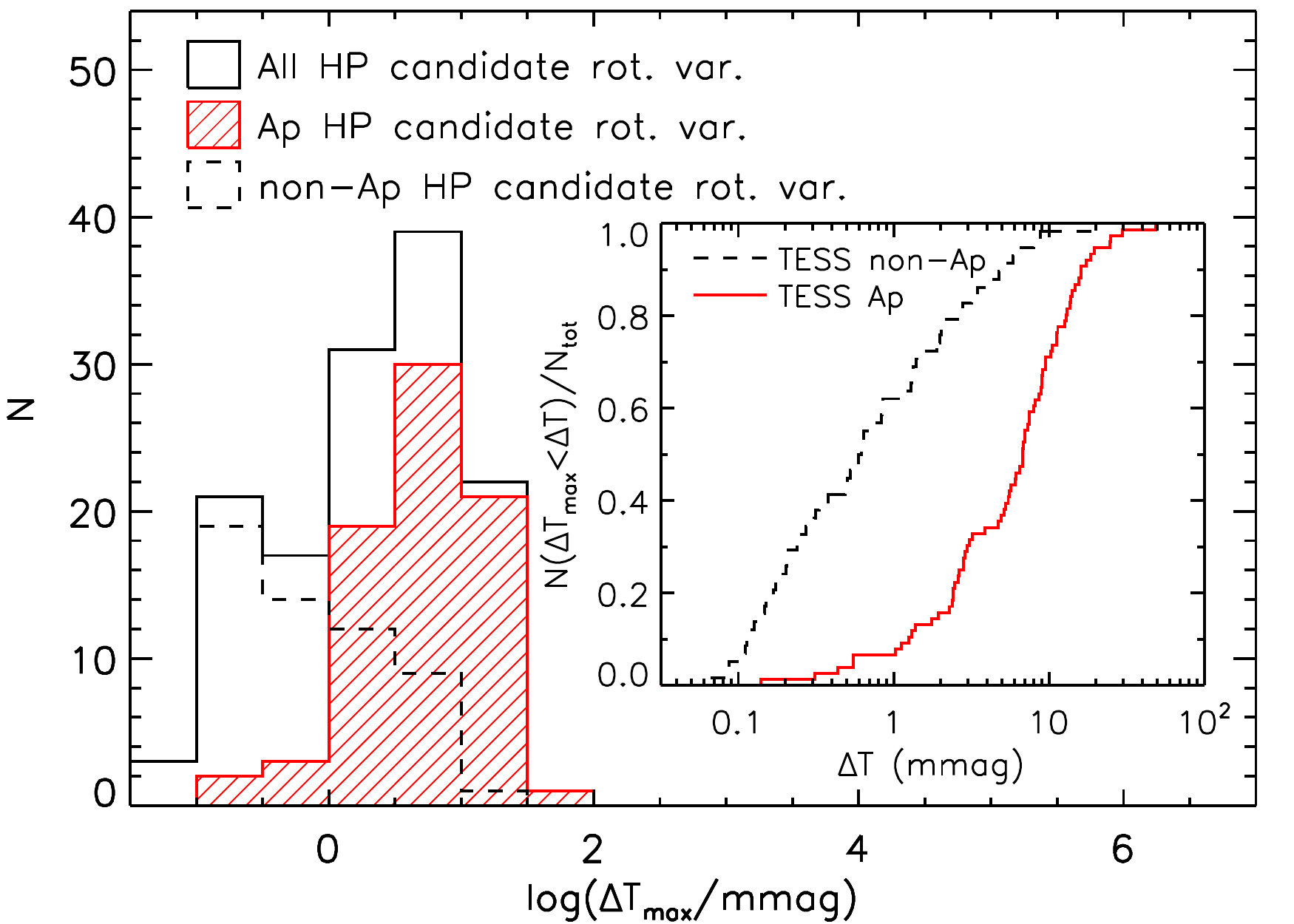}
	\caption{Distributions of the maximum photometric amplitudes ($\Delta T_{\rm max}$) associated with 
	the rotational modulation for the high-probable (HP) rotational variable stars: the total sample 
	(solid black), subsample of Ap stars (red hatched), and subsample of non-Ap stars (dashed black) 
	are shown. The cumulative distribution functions associated with the non-Ap 
	(dashed black) and Ap (solid red) are shown in the inset plot.}
	\label{fig:var_amp}
\end{figure}

\renewcommand{\arraystretch}{1.2}
\begin{table*}
	\caption{Parameters associated with the $134$ identified high probability candidate rotational 
    variables. Columns (1) through (8) list the TIC identifiers, alternative identifiers, spectral 
    types, $V$ magnitudes, maximum photometric amplitudes associated with the rotational modulation 
    signals ($\Delta T_{\rm max}$ where the 1 or 2 subscript indicates whether $\Delta T_{\rm max}$ 
    corresponds to $f_1$ or $f_2$), rotation periods inferred from the \emph{TESS} light curves, 
    published rotation periods, and notes. The numbers in parentheses in rotation periods indicate 
    the uncertainty in the value ($3\sigma$ uncertainties are listed for both $\Delta T_{\rm max}$ and 
    $P_{\rm rot}$). We note the confidence with which any reported magnetic field measurements in the 
    literature have been obtained: definite detections (DD), marginal detections 
    (MD), and null detections (ND). Additionally, we note whether each star is identified as a 
    spectroscopic binary (SB), visual binary (VB), $\delta$~Scuti pulsator, roAp star, and if the 
    amplitude of the rotational modulation is found to vary over time (amplitude modulation, AMod). Those 
    stars without $P_{\rm rot,pub}$ values are considered new rotational variables; similarly, those 
    $\delta$~Scuti and roAp identifications that are not accompanied with references are considered new 
    classifications. The full table appears only in the electronic version of the paper.}
	\label{tbl:vmag_Prot}
	\begin{center}
	\begin{tabular*}{2.0\columnwidth}{@{\extracolsep{\fill}}l c c c c c c r@{\extracolsep{\fill}}}
		\noalign{\vskip-0.2cm}
		\hline
		\hline
		\noalign{\vskip0.5mm}
    TIC  & Alt. ID & Sp. Type & $V$   & $\Delta T_{\rm max}$ & $P_{\rm rot}$  & $P_{\rm rot,pub}$ & Notes \\
         &         &          & (mag) & (mmag)               & (d)            & (d)               &       \\
    (1)  & (2)     & (3)      & (4)   & (5)                  & (6)            & (7)               & (8)   \\
		\noalign{\vskip0.5mm}
		\hline	
		\noalign{\vskip0.5mm}
     7624182 &                 HD~27342 &               A2/3V &       8.8 &$     0.62(2)^1$ &    1.720(1) &                                                   &                                         \\
     7780491 &                 HD~28430 &            ApEuCrSr &       8.2 &$     3.50(2)^2$ &   1.8763(3) &                                                   &                                         \\
    10863314 &                 HD~10653 &                A1IV &       7.7 &$     2.06(2)^2$ &   2.1345(6) &                                                   &                                         \\
    12359289 &                HD~225119 &                ApSi &       8.2 &$     5.14(3)^2$ &   3.0644(5) &            2.944$^{\rm a}$, 3.06395(41)$^{\rm b}$ &                                         \\
    12393823 &                HD~225264 &                A0IV &       8.3 &$     0.83(1)^1$ &   1.4237(6) &                             1.42353(23)$^{\rm b}$ &         SB$^{+}$$^{\rm c}$ ND$^{\rm d}$ \\
    24186142 &                  HD~5601 &                ApSi &       7.6 &$     2.47(3)^1$ &     9.85(2) &                                   1.110$^{\rm e}$ &                            DD$^{\rm f}$ \\
    24225890 &                  HD~5823 &          ApSrEu(Cr) &      10.0 &$     6.88(6)^2$ &    5.007(3) &                                   1.245$^{\rm e}$ &                                         \\
    27985664 &                  HD~3885 &                ApSi &       9.8 &$    11.99(9)^2$ &   1.8144(3) &                                   1.815$^{\rm e}$ &                                         \\
    29432990 &                HD~198966 &                 A9V &       9.2 &$     0.11(2)^1$ &    1.267(3) &                                                   &                                         \\
    29666185 &                HD~199917 &            A2III/IV &       7.1 &$     0.50(1)^1$ &   1.0594(5) &                                                   &                                    AMod \\
    29755072 &                HD~200299 &               A3III &       7.7 &$     0.97(1)^2$ &    6.180(4) &                                                   &                                         \\
    29781099 &                 HD~27997 &            A2mA5-A9 &       8.1 &$    0.392(7)^2$ &   2.8749(9) &                                                   &                                         \\
    32035258 &                 HD~24188 &                ApSi &       6.3 &$    11.40(2)^1$ &   2.2303(1) &             2.230$^{\rm e}$, 2.23047(4)$^{\rm b}$ &                            DD$^{\rm d}$ \\
    38586082 &                 HD~27463 &          ApEuCr(Sr) &       6.3 &$    12.59(3)^1$ &   2.8349(2) &           2.835750$^{\rm g}$, 2.8349(1)$^{\rm b}$ &                $\delta$~Scuti$^{\rm b}$ \\
    41259805 &                 HD~43226 &            ApSr(Eu) &       9.0 &$    11.23(3)^1$ &  1.71450(4) &            1.714$^{\rm e}$, 1.71441(11)$^{\rm b}$ &                          roAp$^{\rm h}$ \\
    42055368 &                 HD~10038 &            A2mA5-F0 &       8.1 &$     3.18(2)^2$ &   2.3120(4) &                                                   &                                         \\
    44627561 &                HD~215559 &                 A7V &       9.3 &$     21.1(1)^1$ &   3.1284(6) &                                1.563140$^{\rm g}$ &                    $\delta$~Scuti, AMod \\
    44678216 &                 HD~25267 &                ApSi &       4.6 &$      7.5(2)^1$ &    3.861(4) &                                   1.210$^{\rm e}$ &                            DD$^{\rm i}$ \\
    44889961 &                 HD~26726 &                ApSr &       9.8 &$    18.66(3)^1$ &   5.3818(9) &                                   5.382$^{\rm e}$ &                                         \\
    52368859 &                 HD~10081 &            ApSr(Eu) &       9.6 &$     9.47(4)^2$ &  1.57052(9) &             1.570$^{\rm e}$, 1.57056(6)$^{\rm b}$ &                                    AMod \\
		\hline \\
    \noalign{\vskip-0.3cm}
\multicolumn{8}{l}{$+$: $P_{\rm orb}=5.400945(40)\,{\rm d}$} \\
\multicolumn{8}{l}{$^{\rm a\,}$\citet{Catalano1998}, $^{\rm b\,}$Cunha et al. 2019 (submitted), $^{\rm c\,}$\citet{Pourbaix2004}, $^{\rm d\,}$\citet{Bagnulo2015}} \\
\multicolumn{8}{l}{$^{\rm e\,}$\citet{Netopil2017}, $^{\rm f\,}$\citet{Kudryavtsev2006}, $^{\rm g\,}$\citet{Oelkers2018}, $^{\rm h\,}$\citet{Borra1980}} \\
	\end{tabular*}
	\end{center}
\end{table*}
\renewcommand{\arraystretch}{1.0}

The distribution of the inferred rotation periods of the $134$ high-probability rotational variables is 
shown in Fig. \ref{fig:Prot_hist}. The longest period is $\approx20\,{\rm d}$; $81$~per~cent of the $134$ 
stars exhibit $P_{\rm rot}<5\,{\rm d}$. Statistically, chemically peculiar Ap stars are known to have 
significantly longer rotation periods compared with their chemically normal, main sequence counterparts 
\citep[e.g.][]{Wolff1975,Abt1995a}. Comparing the high-probability candidate rotational variables 
with Ap classifications with the remainder of the high-probability candidates, we find that the Ap stars 
have statistically longer inferred $P_{\rm rot}$ values: this is apparent from the cumulative distribution 
functions shown in the inset plot of Fig. \ref{fig:Prot_hist}. We also compare the rotation periods of the 
$76$ Ap stars with those of the sample of nearby magnetic Ap stars studied by \citet{Sikora2019a}. The 
\emph{TESS} light curves span a period range of $\approx30\,{\rm d}$ for the majority of the stars in our 
sample (i.e. for those stars that are found in only one sector), which implies that only rotation periods 
of $P_{\rm rot}\lesssim15\,{\rm d}$ can be detected. Comparing the distributions of those stars in the two 
samples having $P_{\rm rot}<15\,{\rm d}$ -- $75/76$ of the Ap stars included in this study and $43/48$ of 
the stars included in the survey carried out by \citet{Sikora2019a} -- yields close agreement.

In Fig. \ref{fig:var_amp} we show the distribution of maximum \emph{TESS} photometric amplitudes 
($\Delta T_{\rm max}$, i.e. the maximum of the amplitudes associated with $f_1$ and $f_2$) for the 
identified $134$ high-probability rotational variables. The distribution ranges from 
$\sim3-52\,{\rm mmag}$ and exhibits a median value $\sim6\,{\rm mmag}$. Comparing the distributions of 
$\Delta T_{\rm max}$ associated with the Ap stars and those not identified as Ap, it is evident that the 
sample of Ap stars tend to exhibit larger $\Delta T_{\rm max}$ values.

We searched the light curves of the identified $134$ high-probability candidate rotational variables for 
evidence of amplitude modulation (AMod). This was carried out by dividing each light curve into sections 
spanning one rotational cycle (defined by $P_{\rm rot}$); those sections having a coverage 
$\lesssim80$~per~cent of the rotational cycle were discarded. The sections were then individually fit 
using Eqn. \ref{eqn:multi_order_fit} with $P$ fixed at $P_{\rm rot}$ while allowing the $a_n$ and 
$\phi_n$ terms to vary. Evidence of AMod is manifest as changes in the amplitudes associated with the 
rotation frequency and first 4 harmonics ($a_n$ where $n$ goes from $1$ to $5$) over time. In many 
cases, the variations in $a_n$ were found to be correlated with the level of noise throughout the light 
curve; those sections of the light curve that were obviously affected by the level of noise were 
disregarded. We identified 22 instances of clear AMod that is not obviously induced by changes in the 
noise level or by the detrending process that was applied to the light curves. These cases are noted in 
Table \ref{tbl:vmag_Prot} and 3 examples of such light curves are shown in Fig. \ref{fig:var_amp}. 
In several of the light curves the detected AMod could be a beating effect caused by low-amplitude 
(pulsation) frequencies in a narrow frequency near the rotation frequency 
\citep[e.g.][]{Degroote2011,Bowman2016} produced by pairs of signals having narrow frequency spacings. 
An example of this is shown in the second panel of Fig. \ref{fig:var_amp} (TIC~70525154). The apparent 
beating could also be induced by the presence of binary companions or background variable stars; 
therefore, the detected AMod may not be intrinsic to the A stars themselves. Out of the 22 stars, 3 have 
likely evolved off of the MS based on their luminosity classes of II or III, 1 is an Am star, 8 are Ap 
stars, and 10 are not identified as chemically peculiar. The inferred rotation periods of the 3 evolved 
stars and 1 Am star are all $<1\,{\rm d}$ while those of the 8 Ap stars and remaining 10 non-Ap stars 
are between $0.5$ and $4.4\,{\rm d}$.

\begin{figure*}
	\centering
	\includegraphics[width=2.0\columnwidth]{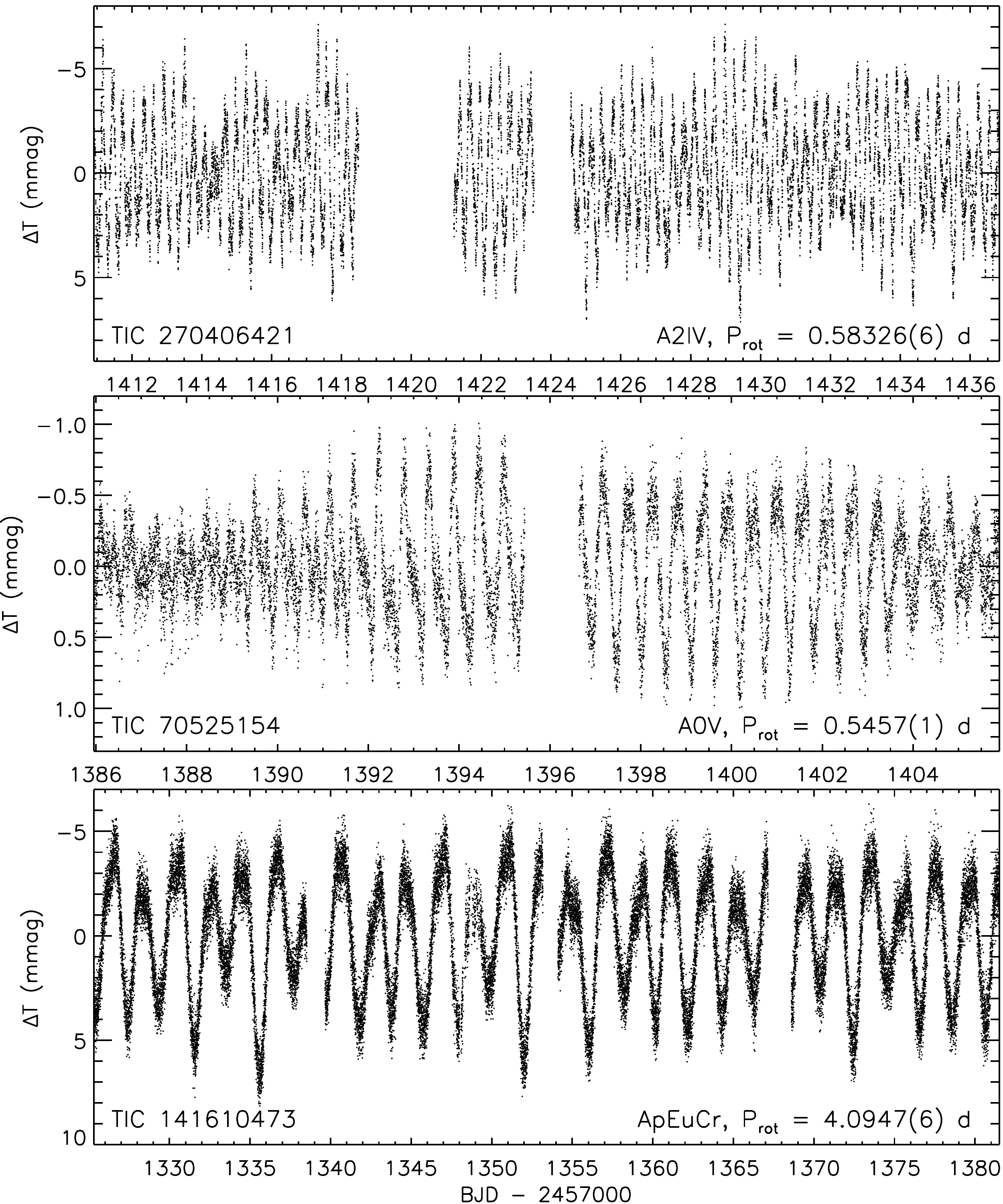}
	\caption{Examples of \emph{TESS} light curves exhibiting rotational modulation with variable 
	amplitudes (amplitude modulation, AMod).}
	\label{fig:var_amp}
\end{figure*}

The study recently carried out by Cunha et al. (submitted) focuses on Ap stars observed with \emph{TESS} 
in sectors 1 and 2. It consists of 83 stars located in sectors 1 and 2; they find that 61 of 
these targets exhibit variability that is consistent with rotational modulation and for which rotation 
periods could be inferred from the \emph{TESS} light curves. A total of 76 out of the 83 stars in their 
sample are also found in our sample while the 7 stars that are excluded have reported spectral types of 
either F or B and thus were filtered out during the construction of our sample. The majority of our 
identifications of candidate rotational variables is in agreement with those reported by Cunha et al. 
(submitted): only four stars that these authors identify as rotational variables are not identified as 
such in our survey due to the difficulty with which we were able to identify clear rotation and first 
harmonic frequency pairs in accordance with the search criteria outlined in Sect. 
\ref{sect:search_criteria}. For example, TIC~394124612 exhibits a large number of low-frequency peaks in 
the periodogram associated with its light curve; as a result, we could not definitively attribute a 
single peak to the star's rotation frequency.

\section{Fundamental parameters}\label{sect:fund_param}

The TIC \citep{Stassun2018} provides a number of fundamental stellar parameters including effective 
temperatures, luminosities, radii, and masses. For $12$ of the $1\,962$ stars in the sample of A-type 
stars the reported fundamental parameters are obtained from large spectroscopic surveys that have been 
compiled into the TIC. For the majority of the stars ($1\,900$ of the $1\,962$ stars in the sample) the 
fundamental parameters listed in the TIC have been derived using the $(V-K_S)-T_{\rm eff}$ calibration 
published by \citet{Huang2015}. Comparisons between the extracted spectral types of the stars in our 
sample with the $T_{\rm eff}$ values derived from this colour-$T_{\rm eff}$ calibration suggest that many 
of the temperatures may be inaccurate (e.g. TIC~13373403 is an A0/1V star with a reported 
$T_{\rm eff}=5\,500\,{\rm K}$ and TIC~30728476 is an A7V star with a reported 
$T_{\rm eff}=9\,200\,{\rm K}$). As a result, we decided to derive the fundamental parameters for our 
sample using several methods.

\subsection{SED fitting}

We derived temperatures, luminosities, and stellar radii by fitting the available photometric 
observations with synthetic spectral energy distributions (SEDs). The TIC contains Johnson $B$ and $V$ 
magnitudes for $1\,956$ of the $1\,962$ A-type stars in our sample and 2MASS $J$, $H$, and $K_S$ 
\citep{Cohen2003} magnitudes for $1\,959$ stars. We searched additional catalogues for available Johnson 
$U$ \citep[$31$ stars,][]{Ducati2002}, Tycho $B_T$ and $V_T$ \citep[$953$ stars,][]{ESA1997}, 
Str{\"o}mgren $uvby$ \citep[$537$ stars,][]{Hauck1997}, and Geneva $UB_1BB_2VV_1G$ 
\citep[$402$ stars,][]{Rufener1988} magnitudes. All of the photometric measurements were converted from 
magnitudes to physical flux units using published Johnson, Tycho \citep{Bessell2012}, Str{\"o}mgren 
\citep{Bessell2011}, Geneva \citep{Rufener1988}, and 2MASS \citep{Cohen2003} zero points. 

Colour excess values ($E[B-V]$) associated with $1\,885$ and $1\,457$ of the $1\,962$ stars in our sample 
are listed in the TIC and reported by \citet{Gontcharov2017}, respectively. We found that de-reddening 
the flux measurements using the values taken from either of these catalogs prior to carrying out the SED 
fitting analysis yielded a large number of $T_{\rm eff}$ values that are significantly greater than is 
expected for A-type stars ($T_{\rm eff}>12\,000\,{\rm K}$). Furthermore, based on the distances 
derived from the Gaia DR2 parallax measurements \citep{Bailer-Jones2018}, many of these stars were found 
to be positioned well below the main sequence in the Hertzsprung-Russell Diagram. As a result, we opted 
to derive the fundamental parameters with $E(B-V)\equiv0$.

\begin{figure*}
	\centering
	\subfigure{\includegraphics[width=2.1\columnwidth]{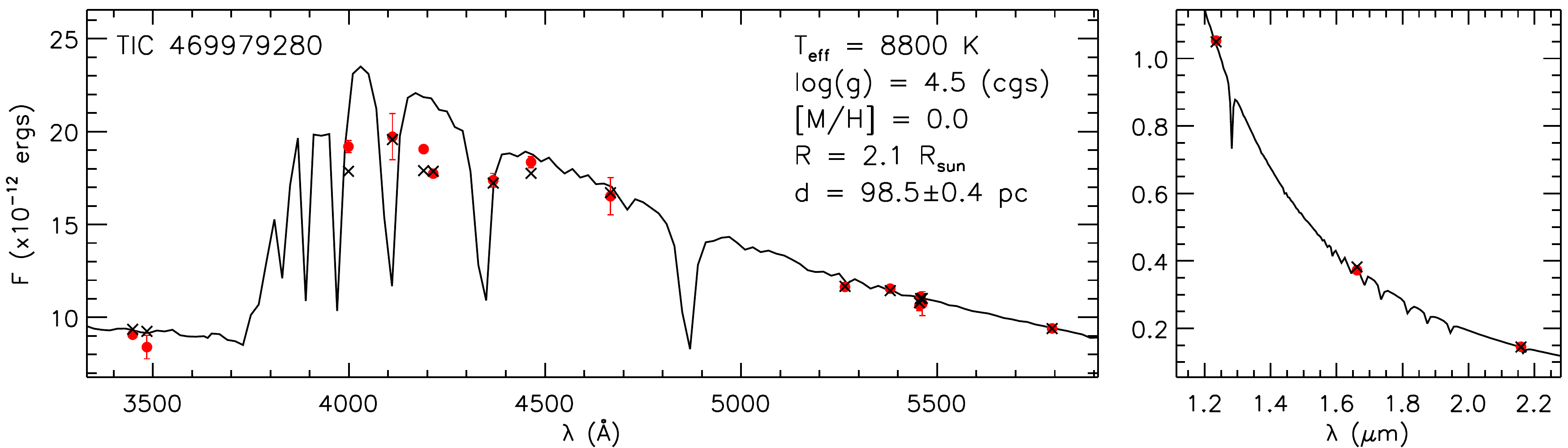}}
	\subfigure{\includegraphics[width=2.1\columnwidth]{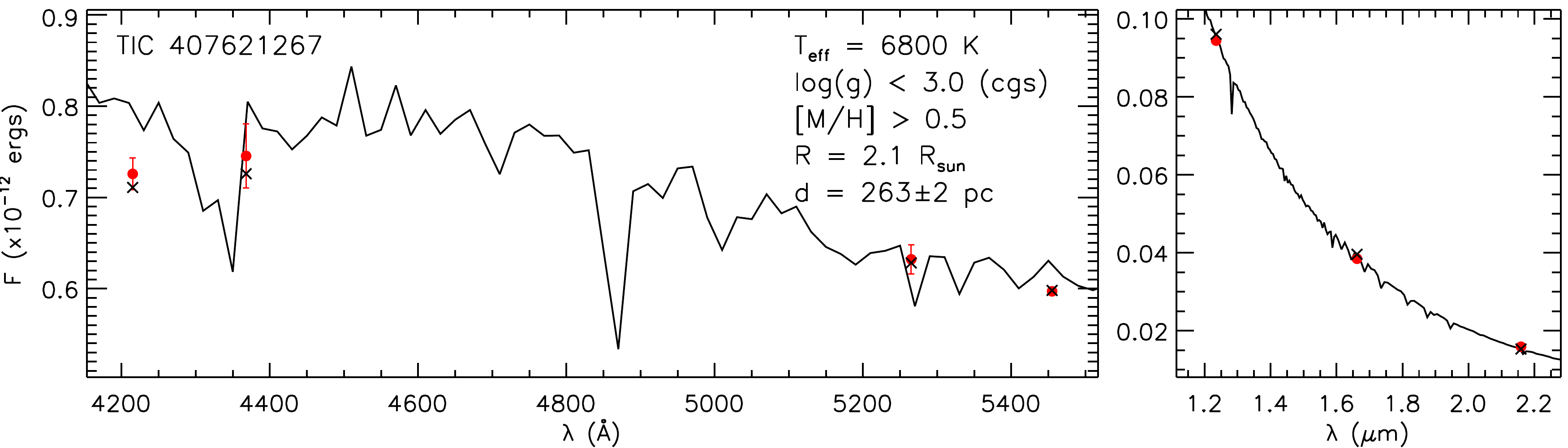}}
	\caption{Two examples of the fits yielded by the SED fitting analysis involving various photometric 
	filters. The filled red circles correspond to the observed flux measurements; the black 
	`X' symbols correspond to the model flux associated with the various filters. \emph{Top:} Fit 
	obtained with Johnson, Tycho, Str{\"o}mgren, Geneva, and 2MASS filters. \emph{Bottom:} Fit 
	obtained with Johnson, Tycho, and 2MASS filters.}
	\label{fig:SED_fit}
\end{figure*}

We used the grid of synthetic SEDs published by \citet{Castelli2004_oth} to fit the photometric 
observations. The grid consists of models having $5\,000\leq T_{\rm eff}\leq20\,000\,{\rm K}$ 
in increments of $250-1\,000\,{\rm K}$, surface gravities of $3\leq\log{g}\leq5\,{\rm (cgs)}$ in 
increments of $0.5$, and metallicities of $-2.5\leq{\rm [M/H]}\leq+0.5$ in increments of $0.2-0.5$. The 
model SEDs are computed using the solar abundances published by \citet{Grevesse1998}. The model flux 
associated with each photometric filter was then computed using the transmission functions obtained from 
the same publications from which the zero points were obtained. The grid of synthetic SEDs was linearly 
interpolated in $T_{\rm eff}$, $\log{g}$, and [M/H] in order to produce a uniform grid having 
$\Delta T_{\rm eff}=50\,{\rm K}$, $\Delta\log{g}=0.5$, and $\Delta{\rm [M/H]}=0.5$.

The best-fitting $T_{\rm eff}$, $\log{g}$, and [M/H] were derived using a grid-search $\chi^2$ 
minimization analysis that was carried out with the interpolated grid of models. The $\chi^2$ value 
associated with each grid point was calculated after deriving an angular radius ($\alpha\equiv (R/d)$, 
where $R$ and $d$ correspond to the star's radius and distance, respectively); for those stars with known 
distances, the angular radius was used to infer the stellar radii. The distances were primarily obtained 
from \citet{Bailer-Jones2018}, which are based on Gaia DR2 parallax measurements ($1\,906$ of $1\,962$ 
stars in the sample). Additional distances were obtained from the Gaia DR1 parallax measurements derived by 
\citet{Astraatmadja2016} ($9$ of $1\,962$ stars), Hipparcos parallax measurements \citep{VanLeeuwen2007} 
($28$ of $1\,962$ stars), or Tycho parallax measurements \citep{ESA1997} ($2$ of $1\,962$ stars); no distances could 
be obtained for $17$ of $1\,962$ stars. For $88$ stars, [M/H] values are listed in the TIC; in these cases, 
[M/H] was fixed at the reported values. The $T_{\rm eff}$ and $R$ values were used to derive the 
luminosities ($L$) of the stars in our sample based on the Stefan-Boltzmann relation. In Fig. 
\ref{fig:SED_fit}, we show two examples of the obtained fits to the observed photometry. The uncertainties 
in $T_{\rm eff}$, $\log{g}$, [M/H], $R$, and $L$ were estimated using the residual bootstrapping method 
described by \citet{Sikora2019}; note that the uncertainties in $R$ and $L$ include the errors in the 
distances that are propagated through the scaling factor $\alpha$.

Effective temperatures were derived for $1\,913$ stars through the SED fitting analysis. Based on our 
bootstrapping method of estimating uncertainties, we find that $1\,765$ of these stars ($92$~per~cent) 
exhibit $\sigma_{T_{\rm eff}}\leq200\,{\rm K}$. For nearly all of those stars that were found to have 
$\sigma_{T_{\rm eff}}\geq400\,{\rm K}$, only Johnson $B$ and $V$ and 2MASS $J$, $H$, and $K_S$ measurements 
are available (only two of the stars with more than five available measurements were found to have 
$\sigma_{T_{\rm eff}}\geq400\,{\rm K}$). Overall, only moderate differences are evident between those 
$\sigma_{T_{\rm eff}}$ values obtained from stars with only five available measurements compared to those 
with more than five: we find that $\langle\sigma_{T_{\rm eff}}\rangle=150\,{\rm K}$ in the former case 
compared to $\langle\sigma_{T_{\rm eff}}\rangle=100\,{\rm K}$ in the latter case.

\subsection{Str{\"o}mgren colour-$T_{\rm eff}$ calibrations}

In addition to the SED fitting, we also derived $T_{\rm eff}$, $R$, and $L$ values using Str{\"o}mgren 
colour-$T_{\rm eff}$ calibrations for those $537$ stars with available Str{\"o}mgren indices. As in the 
SED fitting analysis, we opted not to de-redden the colour indices prior to applying the calibrations.

Out of the $537$ stars with available Str{\"o}mgren indices, $32$ are identified as Ap stars; in these 
cases, we applied the calibrations published by \citet{Stepien1994}. These calibrations include bolometric 
corrections (BCs), which were used along with the available Johnson $V$ magnitudes and distances to derive 
$L$ and $R$. For the remaining stars not identified as Ap, we used the calibrations incorporated into the 
{\sc uvbybeta} IDL routine \citep{Moon1985}. The $R$ and $L$ values were then derived using the same method 
as was applied to the Ap stars where the BCs were calculated from the calibration published by 
\citet{Flower1996}.

\begin{figure}
	\centering
	\includegraphics[width=1.0\columnwidth]{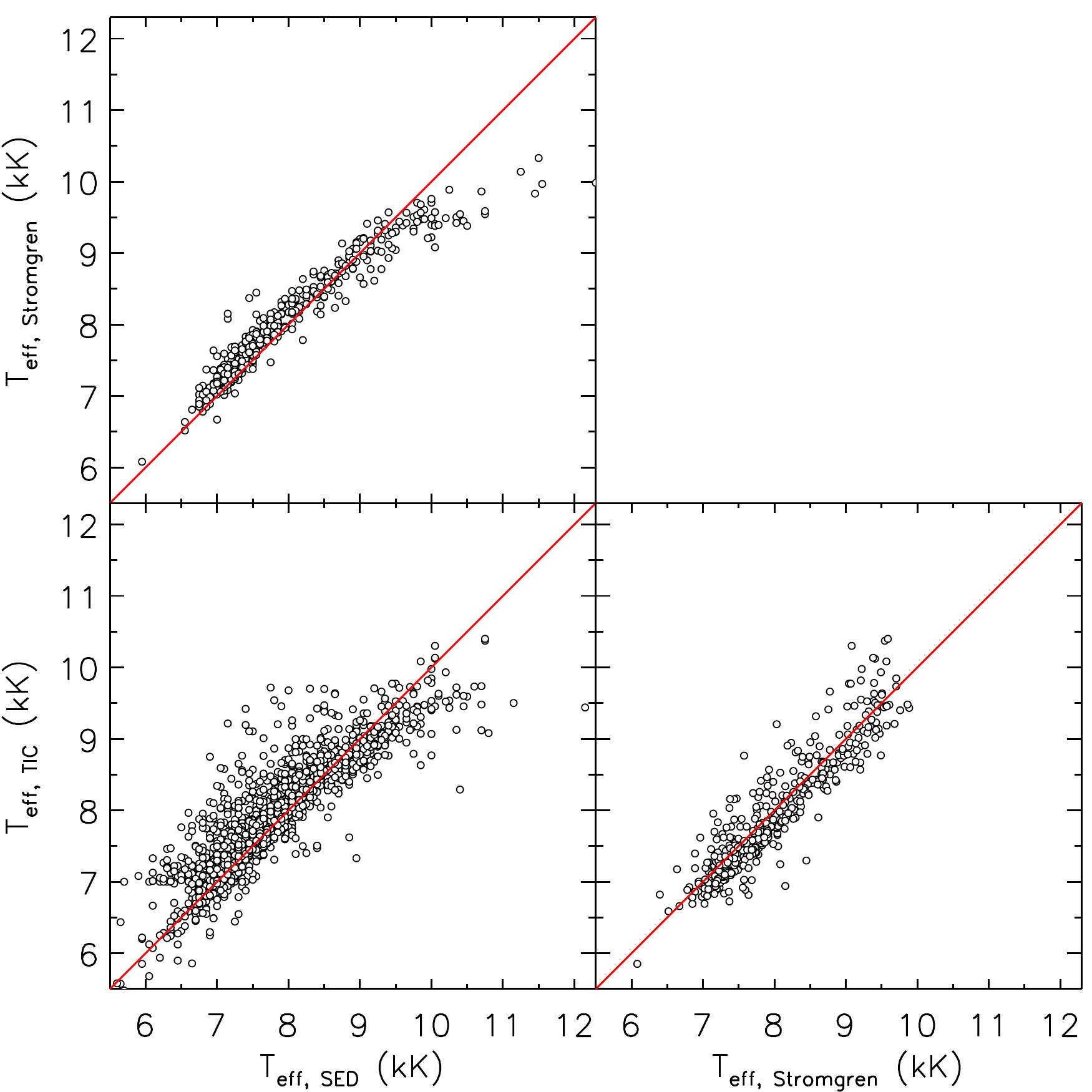}
	\caption{Comparisons between $T_{\rm eff}$ values derived using three methods: $T_{\rm eff, TIC}$ 
	are obtained from the \emph{TESS} Input Catalogue, $T_{\rm eff, Str\ddot{o}mgren}$ are derived from a 
	Str{\"o}mgren colour-$T_{\rm eff}$ calibration, and $T_{\rm eff, SED}$ are derived by fitting 
	synthetic SEDs to various photometric measurements, depending on their availability.}
	\label{fig:T_comp}
\end{figure}

The adopted $T_{\rm eff}$, $R$, and $L$ values associated with the stars in our sample have been derived 
using five methods: (i) a spectroscopic-based analysis \citep{Stassun2018} ($12$ stars), (ii) SED 
modelling ($1\,889$ stars), (iii) Str{\"o}mgren colour-$T_{\rm eff}$ calibrations ($5$ stars) (iv) 
$(V-K_S)-T_{\rm eff}$ calibrations ($41$ stars), and (v) a $(V-B)-T_{\rm eff}$ calibration ($8$ stars). 
For each star, the method that was used to obtain the final parameters was prioritized based on the order 
in which they are listed above. In Fig. \ref{fig:T_comp}, we compare the first three of these methods for 
those stars for which multiple methods could be applied. It is evident that, in general, the $T_{\rm eff}$ 
values derived from the SED modelling and the Str{\"o}mgren colour-$T_{\rm eff}$ calibration are in 
agreement with the values listed in the TIC that were derived from $(V-K_S)-T_{\rm eff}$ calibrations. 
Comparisons between $T_{\rm eff,Str\ddot{o}mgren}$ and $T_{\rm eff,SED}$ yield a median absolute deviation 
of $150\,{\rm K}$ while comparing $T_{\rm eff,Str\ddot{o}mgren}$ and $T_{\rm eff,TIC}$ yields a slightly 
higher value of $210\,{\rm K}$.

\subsection{Hertzsprung-Russell diagram}\label{sect:HRD}

The masses of the stars in our sample were computed by comparing their positions on the HRD with several 
non-rotating grids of model evolutionary tracks. For stars having masses $<3.5\,M_\odot$, we used the 
dense grid computed by \citet{Mowlavi2012}, which has mass intervals of $0.1\,M_\odot$. For the small 
number of more massive stars in the sample, we used the grid computed by \citet{Ekstrom2012}, which has 
larger mass intervals of $0.1-2\,M_\odot$ for the models with $M\leq15\,M_\odot$. The derivation of the 
masses and their uncertainties was carried out using the method described by \citet{Sikora2019}. 

\begin{figure}
	\centering
	\includegraphics[width=1.0\columnwidth]{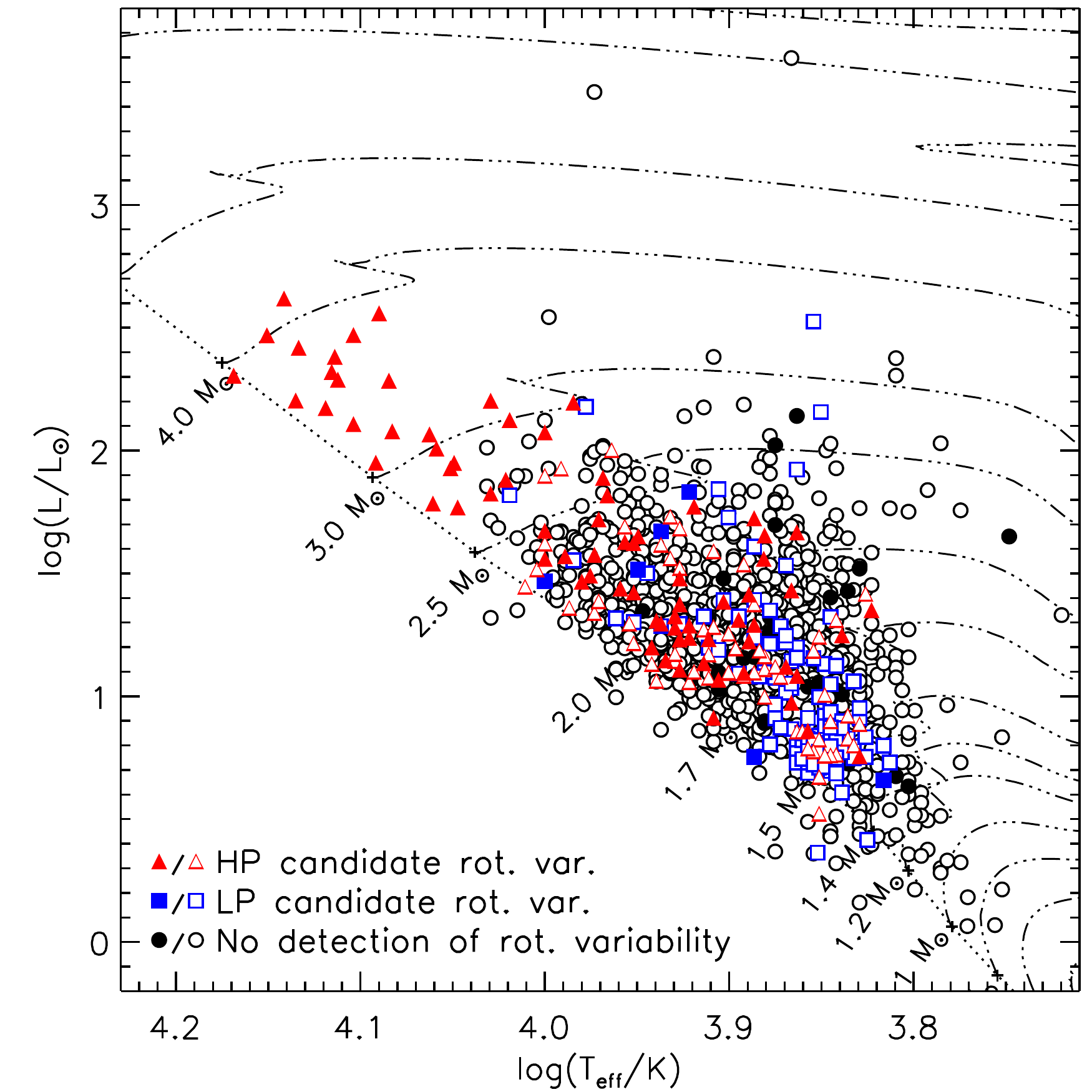}
	\caption{HRD associated with the A-type stars that have been observed with \emph{TESS} in 
	sectors 1 to 4. The different symbols correspond to high-probability (HP) and low-probability (LP) 
	candidate rotational variables (red triangles and blue squares, respectively); black circles correspond 
	to the rest of the sample (i.e. those stars for which variability consistent with rotational 
	modulation was not detected). Filled symbols correspond to Ap stars; open symbols correspond to stars 
	not identified as Ap.}
	\label{fig:hrd}
\end{figure}

\begin{figure}
	\centering
	\includegraphics[width=1.0\columnwidth]{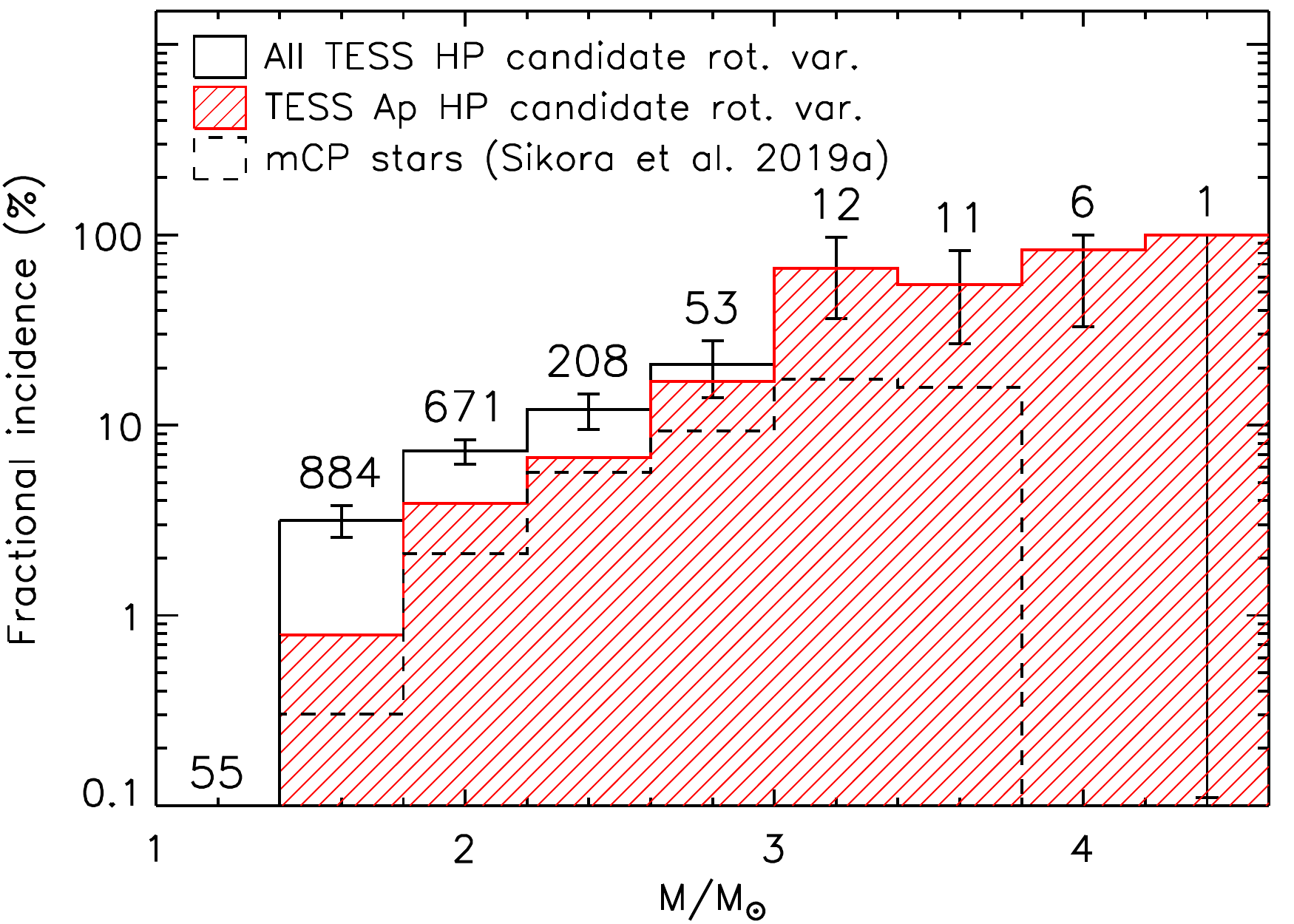}
	\caption{Incidence rate of high-probability (HP) candidate rotational variables as a function of 
	mass where the masses are computed using the derived $T_{\rm eff}$ and $L$ values for the total 
	sample (solid black) and subsample of Ap stars (red hatched). The dashed black line corresponds to 
	the incidence rate of known mCP stars within $100\,{\rm pc}$ \citet{Sikora2019}. The numbers indicate 
	the total number of stars contained in each bin.}
	\label{fig:mass_hist}
\end{figure}

In Fig. \ref{fig:hrd} we show the Hertzsprung-Russell diagram (HRD) generated using the sample's 
derived $T_{\rm eff}$ and $L$ values. These are plotted along with the grid of model evolutionary 
tracks for solar metallicity, non-rotating stars published by \citet{Ekstrom2012}. It is evident that the 
sample roughly spans the entirety of the MS for those stars having $1.4\lesssim M/M_\odot\lesssim3$. Based 
on the HRD, the sample appears to include a small number of post-MS stars, which is consistent with the 
fact that $212$ of the stars in the sample have luminosity classes of II or III. The 
incidence of the high-probability candidate rotational variable stars as a function of mass is shown in 
Fig. \ref{fig:mass_hist}. In Table \ref{tbl:fund_param}, we list the derived fundamental parameters 
($T_{\rm eff}$, $\log{L}$, $M$, and $R$) associated with the $134$ high-probability candidate rotational 
variable stars in our sample.

\renewcommand{\arraystretch}{1.2}
\begin{table*}
	\caption{Fundamental parameters associated with the 134 identified high probability candidate rotational variables. 
      Columns (1) through (7) list the TIC identifiers, distances ($d$), effective temperatures listed in the TIC 
      ($T_{\rm eff,TIC}$), effective temperatures ($T_{\rm eff,SED}$), luminosities ($\log{L_{\rm SED}/L_\odot}$), and 
      radii ($R_{\rm SED}$) derived through the SED fitting analysis, and the masses ($M$) derived from comparisons with 
      evolutionary models. The full table appears only in the elctronic version of the paper.}\label{tbl:fund_param}
	\begin{center}
	\begin{tabular*}{2.0\columnwidth}{@{\extracolsep{\fill}}l c c c c c r@{\extracolsep{\fill}}}
		\noalign{\vskip-0.2cm}
		\hline
		\hline
		\noalign{\vskip0.5mm}
    TIC  & $d$  & $T_{\rm eff,TIC}$ & $T_{\rm eff,SED}$ & $\log(L_{\rm SED}/L_\odot)$ & $R_{\rm SED}$ & $M_{\rm SED}$ \\
         & (pc) & (K)               & (K)               &                             & $(R_\odot)$   & $(M_\odot)$   \\
    (1)  & (2)  & (3)               & (4)               & (5)                         & (6)           & (7)           \\
		\noalign{\vskip0.5mm}
		\hline	
		\noalign{\vskip0.5mm}
     7624182 &      $461\pm7$ &   $8666\pm229$ &   $8450\pm200$ &  $1.68\pm0.04$ &  $3.24\pm0.15$ &  $2.42\pm0.15$ \\
     7780491 &      $218\pm2$ &   $8030\pm220$ &   $7700\pm100$ &  $1.29\pm0.01$ &  $2.48\pm0.05$ &  $1.96\pm0.12$ \\
    10863314 &      $256\pm5$ &   $8968\pm234$ &   $9050\pm100$ &  $1.69\pm0.02$ &  $2.85\pm0.08$ &  $2.45\pm0.14$ \\
    12359289 &     $672\pm31$ &                &  $12300\pm250$ &  $2.56\pm0.07$ &  $4.19\pm0.28$ &  $3.95\pm0.20$ \\
    12393823 &      $226\pm5$ &   $9433\pm241$ &  $10250\pm100$ &  $1.45\pm0.03$ &  $1.68\pm0.05$ &  $2.27\pm0.09$ \\
    24186142 &      $272\pm6$ &                &  $11200\pm150$ &  $1.95\pm0.04$ &  $2.51\pm0.08$ &  $2.94\pm0.18$ \\
    24225890 &      $390\pm8$ &   $7163\pm207$ &   $7300\pm100$ &  $1.08\pm0.02$ &  $2.18\pm0.06$ &  $1.75\pm0.12$ \\
    27985664 &     $872\pm75$ &                &  $11550\pm200$ &  $2.07\pm0.10$ &  $2.69\pm0.27$ &  $3.12\pm0.20$ \\
    29432990 &      $224\pm4$ &   $7115\pm207$ &   $6750\pm100$ &  $0.89\pm0.02$ &  $2.03\pm0.07$ &  $1.58\pm0.11$ \\
    29666185 &      $123\pm1$ &   $7975\pm219$ &   $8200\pm100$ &  $1.27\pm0.02$ &  $2.14\pm0.03$ &  $1.97\pm0.13$ \\
    29755072 &      $232\pm4$ &   $8292\pm224$ &   $7800\pm100$ &  $1.54\pm0.03$ &  $3.21\pm0.09$ &  $2.23\pm0.16$ \\
    29781099 &      $204\pm1$ &   $8362\pm225$ &   $8100\pm100$ &  $1.28\pm0.02$ &  $2.22\pm0.05$ &  $1.97\pm0.12$ \\
    32035258 &      $160\pm2$ &                &  $13650\pm200$ &  $2.20\pm0.03$ &  $2.26\pm0.06$ &  $3.57\pm0.14$ \\
    38586082 &      $126\pm7$ &   $8669\pm229$ &   $9050\pm100$ &  $1.63\pm0.05$ &  $2.65\pm0.19$ &  $2.38\pm0.14$ \\
    41259805 &      $243\pm2$ &   $8034\pm220$ &   $8050\pm100$ &  $1.07\pm0.01$ &  $1.76\pm0.03$ &  $1.79\pm0.11$ \\
    42055368 &      $174\pm1$ &   $7362\pm210$ &   $7600\pm100$ &  $1.11\pm0.01$ &  $2.07\pm0.06$ &  $1.79\pm0.11$ \\
    44627561 &     $408\pm11$ &   $6451\pm198$ &   $6700\pm100$ &  $1.42\pm0.03$ &  $3.79\pm0.18$ &  $1.94\pm0.15$ \\
    44678216 &   $93.6\pm1.9$ &   $12503\pm87$ &  $13000\pm350$ &  $2.38\pm0.04$ &  $3.06\pm0.15$ &  $3.70\pm0.16$ \\
    44889961 &      $442\pm8$ &   $8823\pm232$ &   $8150\pm100$ &  $1.23\pm0.02$ &  $2.07\pm0.07$ &  $1.93\pm0.12$ \\
    52368859 &     $787\pm26$ &   $9076\pm235$ &   $9300\pm100$ &  $1.89\pm0.04$ &  $3.39\pm0.14$ &  $2.72\pm0.16$ \\
		\hline \\
	\end{tabular*}
	\end{center}
\end{table*}
\renewcommand{\arraystretch}{1.0}

\section{Targets of particular interest}\label{sect:targets}

As discussed in Sect. \ref{sect:rot_mod}, we identified $134$ high-probability candidate rotational 
variables. The \emph{TESS} light curves associated with these targets all exhibit low-frequency peaks in 
their LS periodograms that are accompanied by at least one harmonic. Here we discuss some of those targets 
in our sample which are particularly noteworthy.

\subsection{Candidate $\delta$~Scuti and roAp stars}\label{sect:delta_scuti}

Fourteen of the $134$ rotational variable candidates are found to exhibit high-frequency peaks in the 
range $10\lesssim f\lesssim65\,{\rm d^{-1}}$ which are associated with $\delta$~Scuti pulsators 
\citep[e.g.][]{Breger2000b,Holdsworth2014a,Bowman2018} in addition to the low-frequency peaks believed 
to be associated with rotation. Aside from these periodogram features, no other statistically 
significant peaks were detected. Only one of the $14$ stars (TIC~38586082) has been identified as a 
$\delta$~Scuti pulsator in the literature. Little information is available for the other $13$ new 
candidate $\delta$~Scuti stars.

HD~27463 (TIC~38586082) is an ApEuCr(Sr) star \citep{Houk1975} for which a number of photometrically 
determined rotation periods have been reported \citep{Manfroid1981a,Mathys1985}. The $\delta$~Scuti 
pulsations were recently discovered by Cunha et al. (submitted) using the same \emph{TESS} light curves 
analyzed in our study. No magnetic field measurements were found in the literature. Anomalous $\Delta a$ 
measurements of $26\pm6\,{\rm mmag}$ and $28\,{\rm mmag}$ published by \citet{Vogt1998} and 
\citet{Paunzen2005}, respectively, support the star's Ap classification. Speckle interferometry has been 
used to detect a companion with a separation of $0.3\pm0.4\arcsec$ that is dimmer than the primary component 
by $0.9\,{\rm mag}$ in the Str\"omgren $y$ filter \citep{Hartkopf2012}. It is plausible that the 
$\delta$~Scuti pulsation peaks are associated with this dimmer companion rather than the Ap star itself. 
A detailed photometric and spectroscopic analysis of HD~27463 is presented by Khalack et al. (in 
preparation).

In addition to the $\delta$~Scuti stars, our sample of high-probability candidate rotational variable 
stars also contains seven previously discovered roAp stars (TIC~41259805, TIC~211404370, TIC~237336864, 
TIC~326185137, TIC~340006157, TIC~431380369, and TIC~350146296), and one new candidate roAp star 
(TIC~259587315).

HD~30849 (TIC~259587315) is a well-known ApSrCrEu star with a reported rotation period of 
$\sim16\,{\rm d}$ based on Str{\"o}mgren light curves \citep{Renson1979,Hensberge1981}. We infer a 
shorter rotation period of $8.105(6)\,{\rm d}$ based on the \emph{TESS} light curve. \citet{Martinez1994} 
observed this star twice over time spans $\sim1\,{\rm hr}$; no high-overtone pulsations typical of roAp 
stars were detected. Based on the \emph{TESS} light curve, we detect several peaks at frequencies of 
$78.1\,{\rm d^{-1}}$, $80.1\,{\rm d^{-1}}$, and $82.1\,{\rm d^{-1}}$ with respective amplitudes of 
$0.072\,{\rm mmag}$, $0.098\,{\rm mmag}$, and $0.118\,{\rm mmag}$, which suggest that HD~30849 is likely 
a roAp star.

\subsection{Am stars}\label{sect:Am}

\begin{figure*}
	\centering
	\includegraphics[width=2.0\columnwidth]{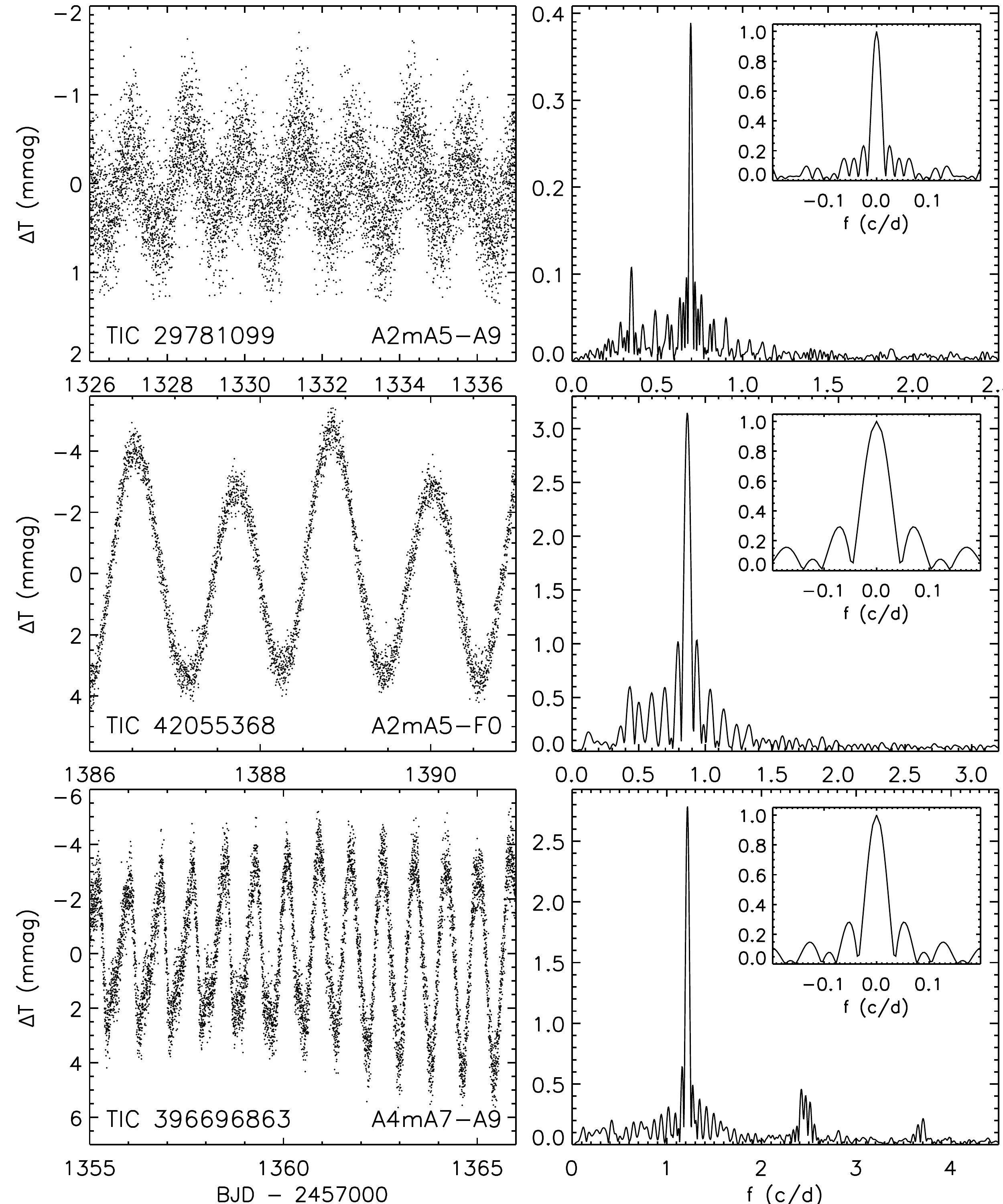}
	\caption{Examples of \emph{TESS} light curves associated with Am stars that are found to exhibit 
	variability that is consistent with a rotational origin. \emph{Left:} Subsamples of the full light 
	curves. \emph{Right:} The Lomb-Scargle periodograms derived from the light curves. The rotational 
	frequencies are apparent along with the first harmonic. The spectral window function is shown in the 
	top right.}
	\label{fig:ex_LC_LS_Am}
\end{figure*}

Three Am stars were identified as being high-probability candidate rotational variables based on our 
analysis of their \emph{TESS} light curves (TIC~29781099, TIC~42055368, and TIC~396696863). All three 
targets are listed in the Catalogue of Ap, HgMn and Am stars compiled by \citet{Renson2009}; however, 
no definitive information could be found in the literature regarding photometric variability, binarity, 
or chemical peculiarities. All three light curves and periodograms are shown in Fig. \ref{fig:ex_LC_LS_Am} 
of the electronic version of the paper.

The long rotation periods associated with Am stars, which allow for the formation of these 
stars' defining chemical peculiarities, are believed to be the result of tidal interactions 
\citep{Abt1961}. This is consistent with the fact that Am stars are commonly found in short period binary 
systems \citep[e.g.][]{Carquillat2007}. Therefore, it is plausible that the periods identified in the 
\emph{TESS} light curves may be associated with orbital motions. In this case, large radial velocity 
variations are expected; spectroscopic monitoring of these targets will need to be carried out in order 
to determine whether these targets are binaries, and that the observed low-frequency harmonics represent 
the orbital period.

\section{Discussion \& Conclusions}\label{sect:disc}

We have identified $134$ high-probability candidate rotational variable stars based on 2~min cadence 
\emph{TESS} light curves among a sample of $1\,962$ A-type stars. More than half of these $134$ stars 
($76$ of $134$) are identified in the literature as Ap stars, which are expected to exhibit photometric 
rotational modulation \citep[e.g.][]{Adelman1992,Catalano1998,GCVS2017,Netopil2017}. Our sample of 
$1\,962$ A-type stars includes $20$ Ap stars for which we did not detect low-frequency variability in 
their \emph{TESS} light curves. These stars are relatively dim compared to the Ap stars for which 
low-frequency variability was detected: $75$~per~cent of the variable Ap stars in our sample have 
brighter $V$ magnitudes than the median magnitude ($V=9.5\,{\rm mag}$) of the $20$ non-variable Ap 
stars. The dimmest variable Ap stars in our sample having $V>9.5\,{\rm mag}$ are found to have 
$\Delta T_{\rm max}>0.3\,{\rm mmag}$; therefore, it is likely that any of the dim, apparently 
non-variable Ap stars that exhibit photometric amplitudes $<0.3\,{\rm mmag}$ fall below our detection 
threshold. That being said, less than $10$~per~cent of the variable Ap stars have 
$\Delta T_{\rm max}<0.3\,{\rm mmag}$, which suggests that only a small number of the $20$ non-variable 
Ap likely have $\Delta T_{\rm max}$ below this limit.

A significant number of Ap stars are known to have very long rotation periods $\gg1\,{\rm yr}$ 
\citep[e.g.][]{Landstreet2000,Mathys2015,Bychkov2016,Mathys2016}. It is possible that several of the 
$20$ apparently non-variable Ap stars are variable over timescales that significantly exceed the 
$28-56\,{\rm d}$ temporal baseline of the \emph{TESS} light curves used in this study. Alternatively, 
assuming that these Ap stars host strong, organized magnetic fields similar to other Ap stars and that 
they have $P_{\rm rot}\lesssim14\,{\rm d}$, the lack of low-frequency variability may either be 
indicative of the geometry and structure of the magnetic field or simply be due to the fact that 
$i=0$~degrees. The Oblique Rotator Model \citep[ORM,][]{Stibbs1950,Preston1967} predicts that no 
longitudinal magnetic field variability will be observed in the case in which (i) the magnetic field is 
axially symmetric and (ii) the axis of symmetry is parallel to the star's axis of rotation. Since the 
chemical spots which produce the non-uniform surface brightness distributions are known to be correlated 
with the magnetic field topology \citep{Kochukhov2004a,Silvester2014a,Kochukhov2017a}, the star's 
rotation might lead to weak or absent photometric variability in this specific case.

The $134$ high-probability candidate rotational variable stars includes $58$ A-type stars that are not 
identified as Ap based on the spectral types obtained from SIMBAD. We find that these stars' light 
curves are statistically distinct from the $76$ identified candidate rotational variable Ap stars both 
in terms of the distribution of inferred $P_{\rm rot}$ values and in terms of their distribution of 
photometric amplitudes, $\Delta T_{\rm max}$. The (presumably) non-Ap stars tend to exhibit much lower 
values of $\Delta T_{\rm max}$ than the Ap stars: $\approx60$~per~cent of the non-Ap stars have 
$\Delta T_{\rm max}\lesssim1\,{\rm mmag}$ compared to $<7$~per~cent of the Ap stars. The distribution of 
rotation periods inferred for the Ap stars in our sample is found to be in agreement with that reported 
by \citet{Sikora2019a} for nearby mCP stars. The non-Ap stars in our sample tend to rotate more rapidly 
based on the inferred rotation periods. This is to be expected from normal (i.e. non-chemically peculiar) 
A-type stars; nevertheless, comparing with published distributions of A star $v\sin{i}$ values 
\citep{Abt1995a,Zorec2012} suggests that the inferred $P_{\rm rot}$ values are unusually long for non-CP 
A stars. Including those non-CP stars identified in our sample as low-probability candidate rotational 
variables reduces the discrepancy but does not change the conclusion that the distribution of periods is 
inconsistent with that of non-CP A-type stars.

The photometric variability associated with Ap stars is well understood to be associated with strong, 
organized, and stable magnetic fields that are visible at the stellar surface. The origin of the 
identified rotational variability associated with the non-Ap star light curves, however, is unclear. 
It is plausible that the A stars are not intrinsically variable but that the observed variability is 
related to late-type binary companions or late-type background stars. A similar explanation for the 
origin of a number of flares in \emph{Kepler} light curves of A-type stars was presented by 
\citet{Pedersen2017}. This is a distinct possibility especially considering the relatively large 
$21\arcsec$ pixel size of the \emph{TESS} CCDs \citep{Sullivan2015} \citep[cf. \emph{Kepler}'s 
$4\arcsec$ pixel size,][]{Koch2010}. 

Assuming that the identified rotational variability of the non-Ap stars is intrinsic to the A stars 
themselves, it cannot be excluded that the variability is caused by the same mechanism that is responsible 
for the formation of dynamic, low-contrast chemical spots on HgMn stars. No convincing evidence of a 
surface magnetism has been found for HgMn stars \citep[e.g.][]{Shorlin2002,Wade2006,Makaganiuk2010}. 
Consequently, it was suggested \citep{Kochukhov2007} that their spots are not linked to magnetic fields 
but are formed under the influence of hydrodynamic instabilities associated with the build-up of 
chemical anomalies by radiative diffusion. Numerical simulations \citep{Alecian2011,Deal2016} 
demonstrate possibility of such instabilities for both late-B and A-type stellar parameter ranges, 
though it is not clear at the moment what governs the horizontal spatial scales of these structures. A 
comparison of the photometric behaviour of known HgMn stars with the present non-Ap sample is necessary 
to assess their similarities and differences. Furthermore, detailed abundance analysis of newly 
discovered rotational variables is required to ascertain their status as chemically normal stars. As 
noted above, the rotation periods associated with our sample of non-Ap stars appear to be unusually long 
for non-CP A stars, which suggests that this subsample may include unrecognised chemically peculiar 
stars with moderate abundance anomalies (e.g. marginal Am and HgMn stars).

The detection of ultra-weak magnetic fields with strengths $\lesssim1\,{\rm G}$ on a small but growing 
number of non-Ap stars such as Vega \citep{Lignieres2009,Petit2011a,Blazere2016} has led to 
speculation that such fields may be widespread amongst the population of non-Ap stars 
\citep[e.g.][]{Petit2011}. Given that (i) chemical and/or temperature spots have been detected on the 
surface of Vega \citep{Bohm2015} and that (ii) these spots exhibit a similarly complex topology to that 
of the detected magnetic field \citep{Petit2010,Petit2017}, it is plausible that the non-Ap candidate 
rotational variables identified in our study also host ultra-weak fields, which are responsible for 
the observed variability. This explanation has been put forth in response to the discovery by 
\citet{Balona2013} that $\sim40$~per~cent of A-type stars observed with \emph{Kepler} appear to 
exhibit rotational variability. \citet{Braithwaite2013b} have proposed that such ultra-weak fields may 
be so-called \emph{failed fossils} -- weak fields having a similar origin to the fossil fields that 
are widely believed to be associated with mCP stars \citep{Cowling1945}. The strengths of failed fossil 
magnetic fields are predicted to decrease as the star evolves across the MS and have strengths that are 
anti-correlated with $P_{\rm rot}$. While \citet{Braithwaite2013b} do not provide any predictions 
regarding photometric variability, the authors note that changes in a failed fossil (and presumably, 
any chemical spots that may be associated with the field) should not be detectable. Evidence of AMod was 
found in the light curves of 10 of the 48 high-probability candidate rotational variable non-CP MS 
stars in our sample, which could be indicative of evolving surface spot morphologies. Assuming that the 
detected AMod is intrinsic to the A stars and not the result of contamination from binary companions or 
background stars, these detections would be inconsistent with the failed fossil model.

\citet{Cantiello2019} explored an alternative explanation for the origin of ultra-weak fields in 
A- and late B-type stars, which involves dynamo fields that are generated near the stellar 
surface. The authors suggest that (i) all rapidly rotating A- and B-type stars possess sub-surface 
H and He convection zones where turbulent dynamo may generate magnetic fields and that (ii) these fields 
are associated with bright temperature spots similar to those predicted to exist on more massive stars 
\citep{Cantiello2011}. A rotation-activity connection is known to be a key feature of any type of 
dynamo-powered magnetic field and associated surface activity. According to \citet{Cantiello2019}, the 
dynamo action requires stellar rotation with periods of $0.5-1\,{\rm d}$ for a dwarf star with 
parameters similar to Vega. Among the 48 MS non-CP stars in our sample only $20$ have 
$P_{\rm rot}\leq1.5\,{\rm d}$ and the rest are slower rotators with $P_{\rm rot}$ up to $7.5\,{\rm d}$ 
for which no dynamo action is anticipated. Moreover, we do not observe a clear anticorrelation between 
$P_{\rm rot}$ and $\Delta T_{\rm max}$, which is expected within the framework of any dynamo hypothesis. 
As noted above, we do find evidence of AMod in the light curves of 10 of the 48 MS non-CP stars. These 
10 stars have inferred rotation periods ranging from $0.5$ to $4.6\,{\rm d}$ and all but 4 have 
$P_{\rm rot}\leq1.5\,{\rm d}$. We also note that we did not find evidence of AMod for an additional 17 
of the MS non-CP stars with $P_{\rm rot}\leq1.5\,{\rm d}$.

It is likely that in order to better understand the origin of the variability associated with the 
non-Ap stars identified in our sample as being rotational modulation we will need to obtain 
high-resolution spectroscopic measurements. The measurement of chemical abundancies in the atmosphere's 
of these stars is of particular interest as it may confirm our suspicion that some of the stars in our 
sample are at least weakly chemically peculiar. Detecting spectroscopic variability or placing 
constraints on the strength of such variability would also be useful in this regard. We also note that 
combining the \emph{TESS} light curves with multi-colour long-term monitoring, such as that provided by 
the \emph{BRITE}-Constellation mission \citep{Weiss2008}, may provide insight into the puzzling results 
presented here.

\section*{Acknowledgments}

GAW acknowledges support in the form of a Discovery Grant from the Natural Science and Engineering 
Research Council (NSERC) of Canada. ADU and VK acknowledge support from NSERC. The research leading 
to these results has received funding from the European Research Council (ERC) under the European 
Union’s Horizon 2020 research and innovation programme (grant agreement No. 670519: MAMSIE). SC 
gratefully acknowledge funding through grant 2015/18/A/ST9/00578 of the Polish National Science Centre 
(NCN). OK acknowledges support by the Swedish Research Council (project 621-2014-5720) and the Swedish 
National Space Board (projects 185/14, 137/17).

This research has made use of the SIMBAD database, operated at CDS, Strasbourg, France. This paper includes 
data collected with the TESS mission, obtained from the MAST data archive at the Space Telescope Science 
Institute (STScI). Funding for the TESS mission is provided by the NASA Explorer Program. STScI is operated 
by the Association of Universities for Research in Astronomy, Inc., under NASA contract NAS 5–26555. 
Funding for the TESS Asteroseismic Science Operations Centre is provided by the Danish National Research 
Foundation (Grant agreement no.: DNRF106), ESA PRODEX (PEA 4000119301) and Stellar Astrophysics Centre 
(SAC) at Aarhus University. We thank the TESS team and staff and TASC/TASOC for their support of
the present work. We thank Dr.~D.~W.~Kurtz for providing useful comments.

\bibliography{jsikora_tess_rot_var_Astars}
\bibliographystyle{mn2e}

\section*{Appendix}

\clearpage

\onecolumn
\tablecaption{Parameters associated with the $134$ identified high probability candidate rotational 
    variables. Columns (1) through (8) list the TIC identifiers, alternative identifiers, spectral 
    types, $V$ magnitudes, maximum photometric amplitudes associated with the rotational modulation 
    signals ($\Delta T_{\rm max}$ where the 1 or 2 subscript indicates whether $\Delta T_{\rm max}$ 
    corresponds to $f_1$ or $f_2$), rotation periods inferred from the \emph{TESS} light curves, 
    published rotation periods, and notes. The numbers in parentheses in rotation periods indicate 
    the uncertainty in the value ($3\sigma$ uncertainties are listed for both $\Delta T_{\rm max}$ and 
    $P_{\rm rot}$). We note the confidence with which any reported magnetic field measurements in the 
    literature have been obtained: definite detections (DD), marginal detections 
    (MD), and null detections (ND). Additionally, we note whether each star is identified as a 
    spectroscopic binary (SB), visual binary (VB), $\delta$~Scuti pulsator, roAp star, and if the 
    amplitudes of the rotational modulation is found to vary over time (amplitude modulation, AMod). Those 
    stars without $P_{\rm rot,pub}$ values are considered new rotational variables; similarly, those 
    $\delta$~Scuti and roAp identifications that are not accompanied with references are considered new 
    classifications. The full table appears only in the electronic version of the 
    paper.}\label{tbl:vmag_Prot_full}
\tablefirsthead{
    \hline
    \hline
    \noalign{\vskip0.5mm}
    TIC  & Alt. ID & Sp. Type & $V$   & $\Delta T_{\rm max}$ & $P_{\rm rot}$  & $P_{\rm rot,pub}$ & Notes \\
         &         &          & (mag) & (mmag)               & (d)            & (d)               &       \\
    (1)  & (2)     & (3)      & (4)   & (5)                  & (6)            & (7)               & (8)   \\
    \noalign{\vskip0.5mm}
    \hline
    \noalign{\vskip0.5mm}
}
\tablehead{
    \multicolumn{8}{l}{continued from previous page}\\
    \hline
    \hline
    \noalign{\vskip0.5mm}
    TIC  & Alt. ID & Sp. Type & $V$   & $\Delta T_{\rm max}$ & $P_{\rm rot}$  & $P_{\rm rot,pub}$ & Notes \\
         &         &          & (mag) & (mmag)               & (d)            & (d)               &       \\
    (1)  & (2)     & (3)      & (4)   & (5)                  & (6)            & (7)               & (8)   \\
    \noalign{\vskip0.5mm}
    \hline  
    \noalign{\vskip0.5mm}
}
\tabletail{
    \noalign{\vskip0.5mm}\hline
    \multicolumn{8}{l}{continued on next page}\\
}
\tablelasttail{
    \noalign{\vskip0.5mm}\hline
\multicolumn{8}{l}{$+$: $P_{\rm orb}=5.400945(40)\,{\rm d}$, $++$: $P_{\rm orb}=464.66\,{\rm yrs}$} \\
\multicolumn{8}{l}{$^{\rm a\,}$\citet{Catalano1998}, $^{\rm b\,}$Cunha et al. 2019 (submitted), $^{\rm c\,}$\citet{Pourbaix2004}, $^{\rm d\,}$\citet{Bagnulo2015}} \\
\multicolumn{8}{l}{$^{\rm e\,}$\citet{Netopil2017}, $^{\rm f\,}$\citet{Kudryavtsev2006}, $^{\rm g\,}$\citet{Oelkers2018}, $^{\rm h\,}$\citet{Borra1980}} \\
\multicolumn{8}{l}{$^{\rm i\,}$\citet{Malkov2012}, $^{\rm j\,}$\citet{Sikora2019}, $^{\rm k\,}$\citet{Kurtz1984}, $^{\rm l\,}$\citet{Bohlender1993}} \\
\multicolumn{8}{l}{$^{\rm m\,}$\citet{Martinez1990}, $^{\rm n\,}$\citet{Mathys1997a}, $^{\rm o\,}$\citet{Borra1975}, $^{\rm p\,}$\citet{Maitzen1980}} \\
\multicolumn{8}{l}{$^{\rm q\,}$\citet{Auriere2007}, $^{\rm r\,}$\citet{Kurtz1985}, $^{\rm s\,}$\citet{Kurtz1990a}, $^{\rm t\,}$\citet{Manfroid1983b}} \\
}
\centering
\begin{mpsupertabular*}{1.0\columnwidth}{@{\extracolsep{\fill}}l c c c c c c r@{\extracolsep{\fill}}}
     7624182 &                 HD~27342 &               A2/3V &       8.8 &$     0.62(2)^1$ &    1.720(1) &                                                   &                                         \\
     7780491 &                 HD~28430 &            ApEuCrSr &       8.2 &$     3.50(2)^2$ &   1.8763(3) &                                                   &                                         \\
    10863314 &                 HD~10653 &                A1IV &       7.7 &$     2.06(2)^2$ &   2.1345(6) &                                                   &                                         \\
    12359289 &                HD~225119 &                ApSi &       8.2 &$     5.14(3)^2$ &   3.0644(5) &            2.944$^{\rm a}$, 3.06395(41)$^{\rm b}$ &                                         \\
    12393823 &                HD~225264 &                A0IV &       8.3 &$     0.83(1)^1$ &   1.4237(6) &                             1.42353(23)$^{\rm b}$ &         SB$^{+}$$^{\rm c}$ ND$^{\rm d}$ \\
    24186142 &                  HD~5601 &                ApSi &       7.6 &$     2.47(3)^1$ &     9.85(2) &                                   1.110$^{\rm e}$ &                            DD$^{\rm f}$ \\
    24225890 &                  HD~5823 &          ApSrEu(Cr) &      10.0 &$     6.88(6)^2$ &    5.007(3) &                                   1.245$^{\rm e}$ &                                         \\
    27985664 &                  HD~3885 &                ApSi &       9.8 &$    11.99(9)^2$ &   1.8144(3) &                                   1.815$^{\rm e}$ &                                         \\
    29432990 &                HD~198966 &                 A9V &       9.2 &$     0.11(2)^1$ &    1.267(3) &                                                   &                                         \\
    29666185 &                HD~199917 &            A2III/IV &       7.1 &$     0.50(1)^1$ &   1.0594(5) &                                                   &                                    AMod \\
    29755072 &                HD~200299 &               A3III &       7.7 &$     0.97(1)^2$ &    6.180(4) &                                                   &                                         \\
    29781099 &                 HD~27997 &            A2mA5-A9 &       8.1 &$    0.392(7)^2$ &   2.8749(9) &                                                   &                                         \\
    32035258 &                 HD~24188 &                ApSi &       6.3 &$    11.40(2)^1$ &   2.2303(1) &             2.230$^{\rm e}$, 2.23047(4)$^{\rm b}$ &                            DD$^{\rm d}$ \\
    38586082 &                 HD~27463 &          ApEuCr(Sr) &       6.3 &$    12.59(3)^1$ &   2.8349(2) &           2.835750$^{\rm g}$, 2.8349(1)$^{\rm b}$ &                $\delta$~Scuti$^{\rm b}$ \\
    41259805 &                 HD~43226 &            ApSr(Eu) &       9.0 &$    11.23(3)^1$ &  1.71450(4) &            1.714$^{\rm e}$, 1.71441(11)$^{\rm b}$ &                          roAp$^{\rm h}$ \\
    42055368 &                 HD~10038 &            A2mA5-F0 &       8.1 &$     3.18(2)^2$ &   2.3120(4) &                                                   &                                         \\
    44627561 &                HD~215559 &                 A7V &       9.3 &$     21.1(1)^1$ &   3.1284(6) &                                1.563140$^{\rm g}$ &                    $\delta$~Scuti, AMod \\
    44678216 &                 HD~25267 &                ApSi &       4.6 &$      7.5(2)^1$ &    3.861(4) &                                   1.210$^{\rm e}$ &                            DD$^{\rm i}$ \\
    44889961 &                 HD~26726 &                ApSr &       9.8 &$    18.66(3)^1$ &   5.3818(9) &                                   5.382$^{\rm e}$ &                                         \\
    52368859 &                 HD~10081 &            ApSr(Eu) &       9.6 &$     9.47(4)^2$ &  1.57052(9) &             1.570$^{\rm e}$, 1.57056(6)$^{\rm b}$ &                                    AMod \\
    55400261 &                 HD~30296 &                 A9V &       8.7 &$     1.08(1)^1$ &    4.606(2) &                                                   &                    $\delta$~Scuti, AMod \\
    66646031 &                 HD~16145 &            ApCrEuSr &       7.7 &$    13.65(3)^2$ &   4.4751(4) &                                   2.238$^{\rm e}$ &                                         \\
    67650835 &                  HD~7676 &            ApSrCrEu &       8.4 &$    14.72(2)^1$ &   5.0979(5) &                                   5.098$^{\rm e}$ &                                         \\
    70525154 &                 HD~13709 &                 A0V &       5.3 &$    0.332(8)^1$ &   0.5457(1) &                                                   &                                    AMod \\
    79272047 &                HD~201801 &                 A9V &       8.8 &$     0.20(2)^1$ &     5.23(3) &                                                   &                                         \\
    79394646 &         TYC 8793-01478-1 &              A5IV-V &       4.4 &$     1.48(1)^1$ &    4.132(4) &                                                   &                    $\delta$~Scuti, AMod \\
    89545031 &                HD~223640 &            A0VpSiSr &       5.2 &$     9.06(2)^1$ &   3.7342(5) &            3.735$^{\rm e}$, 3.72251(97)$^{\rm b}$ &                            ND$^{\rm d}$ \\
    92705248 &                HD~200623 &            ApSrEuCr &       9.1 &$     9.49(4)^1$ &   2.1576(2) &              2.200$^{\rm e}$, 2.1577(2)$^{\rm b}$ &                                         \\
   102090493 &                  HD~7454 &                A5IV &       9.5 &$      8.8(1)^2$ &   1.4585(2) &                                0.729208$^{\rm g}$ &                          $\delta$~Scuti \\
   129636548 &                HD~203585 &                ApSi &       5.8 &$    1.385(7)^1$ &   3.1082(5) &                             3.11016(56)$^{\rm b}$ &                     VB$^{++}$$^{\rm j}$ \\
   136843852 &                  HD~9335 &             A5/7III &       7.5 &$     4.12(1)^2$ &   1.8595(1) &                                                   &                                         \\
   140044682 &                HD~208489 &                 A0V &       8.7 &$     5.85(2)^1$ &  0.92045(5) &                                                   &                                         \\
   140204398 &                 HD~24825 &          ApCrEu(Sr) &       6.8 &$    11.25(2)^1$ &    6.795(1) &                                                   &                                         \\
   141028198 &                 HD~35361 &              ApCrEu &       9.9 &$     5.06(3)^2$ &    6.306(1) &                               6.3035(9)$^{\rm b}$ &                                         \\
   141610473 &                 HD~41613 &              ApEuCr &       9.7 &$     3.08(3)^2$ &   4.0947(6) &                               4.0954(4)$^{\rm b}$ &                                    AMod \\
   144069014 &                HD~213230 &               A5/6V &       7.7 &$     0.49(1)^1$ &   1.3559(7) &                                                   &                          $\delta$~Scuti \\
   147086189 &                HD~203898 &                 A9V &       9.3 &$     1.34(2)^1$ &   1.0394(3) &                                                   &                    $\delta$~Scuti, AMod \\
   150250959 &                 HD~44532 &                 A2V &       8.8 &$     0.52(2)^1$ &    2.859(5) &                                                   &                                         \\
   153742460 &                 HD~28299 &                ApSi &       7.6 &$     4.94(2)^2$ &   3.3639(4) &                                                   &                                         \\
   155945483 &                  HD~1948 &               A2/3V &       8.1 &$     0.14(1)^1$ &     6.10(3) &                                                   &                                         \\
   159834975 &                HD~203006 &            ApCrEuSr &       4.8 &$    5.469(8)^2$ &  2.12199(6) &             2.122$^{\rm e}$, 2.12230(9)$^{\rm b}$ &                            DD$^{\rm k}$ \\
   161270578 &                HD~215789 &               A2IVn &       3.5 &$    0.088(4)^1$ &   0.7910(7) &                                                   &                            ND$^{\rm d}$ \\
   161334416 &                HD~216485 &                 A1V &       8.9 &$     1.90(2)^1$ &   1.5264(4) &                                                   &                                         \\
   161570223 &                 HD~31973 &            ApSrCrEu &       9.4 &$     1.19(3)^2$ &    2.705(1) &                                                   &                                         \\
   182909257 &                  HD~6783 &                ApSi &       8.0 &$     2.87(2)^1$ &    3.137(1) &                             3.14108(82)$^{\rm b}$ &                                         \\
   183802606 &                  HD~8700 &                ApSi &       9.6 &$     2.64(2)^1$ &   2.2701(3) &                             2.27015(17)$^{\rm b}$ &                                         \\
   183802904 &                  HD~8783 &            ApSrCrEu &       7.8 &$    10.30(1)^1$ &   19.384(7) &            19.396$^{\rm e}$, 19.408(17)$^{\rm b}$ &                            ND$^{\rm d}$ \\
   200441087 &                 HD~30335 &            ApSrCrEu &       9.7 &$     7.35(6)^1$ &    5.094(2) &               5.100$^{\rm e}$, 5.096320$^{\rm g}$ &                                         \\
   200783972 &                 HD~21360 &                 A0V &       6.5 &$    0.272(9)^1$ &    1.786(2) &                               27.151800$^{\rm g}$ &                                         \\
   201923258 &                 HD~17450 &              A0IV/V &       8.9 &$     0.84(2)^1$ &    2.396(2) &                                                   &                                         \\
   204293493 &                HD~217448 &           A8V+(G/K) &       9.6 &$     0.17(2)^1$ &    2.162(7) &                                                   &                                         \\
   204314449 &         TYC 6397-01365-2 &                   ~ &       6.8 &$     0.34(1)^2$ &   1.5095(5) &                                                   &                                    AMod \\
   206648435 &                HD~215983 &            ApSrEuCr &       9.7 &$     8.89(3)^1$ &    5.157(2) &                              5.1094(22)$^{\rm b}$ &                                         \\
   206663039 &                HD~216336 &          A0VpSrCrEu &       4.5 &$    0.548(7)^1$ &    2.542(1) &                                                   &                                         \\
   207143419 &                 HD~18796 &                 A1V &       8.8 &$     0.64(2)^1$ &    1.900(2) &                                                   &                                    AMod \\
   207208753 &                 HD~20505 &              ApCrSr &       9.8 &$     9.01(5)^1$ &   2.0436(4) &            2.044$^{\rm e}$, 2.04334(19)$^{\rm b}$ &                                         \\
   211404370 &                HD~203932 &              ApSrEu &       8.8 &$     0.55(2)^1$ &     6.64(3) &                               6.442(12)$^{\rm b}$ &             roAp$^{\rm l}$ MD$^{\rm d}$ \\
   219118940 &                HD~214582 &                A2II &       9.5 &$     7.99(8)^1$ &   0.7927(1) &                                0.792111$^{\rm g}$ &                    $\delta$~Scuti, AMod \\
   219234021 &                 HD~27211 &            ApSrCrEu &       9.4 &$     3.23(3)^1$ &   1.1634(2) &                                1.163440$^{\rm g}$ &                                         \\
   219990936 &                 HD~12671 &               A8III &       8.5 &$     2.02(3)^1$ &   0.7140(1) &                                                   &                    $\delta$~Scuti, AMod \\
   220035931 &                 HD~34631 &                ApSi &       7.0 &$    19.66(4)^1$ &  2.20280(9) &               1.822620$^{\rm g}$, 2.203$^{\rm e}$ &                                         \\
   220414891 &                 HD~30609 &              A2IV/V &       8.8 &$     0.23(1)^2$ &     6.92(1) &                                                   &                                         \\
   220565429 &                 HD~19398 &            A9III/IV &       8.8 &$     0.21(1)^1$ &   0.5566(3) &                                                   &                                         \\
   220570020 &                 HD~19695 &                 A9V &       9.4 &$     0.09(1)^1$ &    1.345(3) &                                                   &                                         \\
   231813751 &                 HD~38471 &                ApSi &       7.6 &$     5.59(9)^1$ &   2.4192(9) &                                                   &                                    AMod \\
   231844926 &                 HD~10840 &                ApSi &       6.8 &$   24.863(6)^1$ & 2.097679(7) &              2.098$^{\rm e}$, 2.0971(1)$^{\rm b}$ &                            MD$^{\rm d}$ \\
   231864288 &                  HD~1860 &                 A0V &       7.3 &$    0.127(8)^1$ &   1.4968(9) &              32.372898$^{\rm g}$, 8.634$^{\rm e}$ &                                         \\
   232066526 &                 HD~11090 &                ApSr &      10.8 &$     6.74(4)^2$ &   2.9195(2) &                             2.91982(16)$^{\rm b}$ &                                         \\
   234346165 &                 HD~16504 &                ApSi &       9.1 &$     6.14(4)^1$ &   3.3041(4) &                               3.3040(3)$^{\rm b}$ &                                         \\
   235007556 &                HD~221006 &                ApSi &       5.7 &$    15.62(1)^1$ &  2.31475(5) &        2.31206(36)$^{\rm b}$, 27.151800$^{\rm g}$ &                            DD$^{\rm m}$ \\
   237336864 &                HD~218495 &              ApEuSr &       9.4 &$    12.92(2)^2$ &   4.2006(3) &                               4.2006(1)$^{\rm b}$ &             roAp$^{\rm n}$ DD$^{\rm d}$ \\
   259587315 &                 HD~30849 &            ApSrCrEu &       8.9 &$     4.72(3)^1$ &    7.489(8) &                                  15.864$^{\rm e}$ &                                    roAp \\
   262613883 &                 HD~63728 &              ApEuCr &       9.4 &$    24.78(6)^1$ &   1.8402(1) &            1.84015(17)$^{\rm b}$, 1.840$^{\rm e}$ &                                         \\
   269857621 &                 HD~31230 &                 A1V &       8.6 &$     4.15(1)^1$ &  1.12324(2) &                                                   &                                         \\
   270250508 &                HD~209133 &                 A9V &       8.3 &$     0.15(1)^1$ &   0.7901(4) &                                                   &                                         \\
   270304671 &                HD~209605 &            ApSrEuCr &       9.6 &$    13.09(3)^1$ &    7.822(2) &             7.8896(50)$^{\rm b}$, 7.813$^{\rm e}$ &                                         \\
   270406421 &                 HD~13467 &                A2IV &       6.7 &$     2.64(5)^2$ &  0.58326(6) &                                                   &                    $\delta$~Scuti, AMod \\
   272316843 &                 HD~66082 &               A0/1V &       9.3 &$     0.25(2)^1$ &    1.265(2) &                                                   &                                         \\
   277688819 &                HD~208217 &            ApSrEuCr &       7.2 &$    12.89(1)^1$ &   8.4464(9) &             8.3200(84)$^{\rm b}$, 8.445$^{\rm e}$ &                            DD$^{\rm o}$ \\
   277748932 &                HD~208759 &            ApSrEuCr &      10.0 &$     2.38(4)^2$ &    4.452(3) &                              4.4501(19)$^{\rm b}$ &                                         \\
   278804454 &                HD~212385 &            ApSrEuCr &       6.8 &$     6.99(1)^1$ &   2.5084(2) &              2.5062(2)$^{\rm b}$, 2.480$^{\rm e}$ &                            DD$^{\rm d}$ \\
   279573219 &                 HD~54118 &                ApSi &       5.1 &$     8.01(3)^1$ &   3.2743(4) &              27.151800$^{\rm g}$, 3.275$^{\rm e}$ &                            DD$^{\rm p}$ \\
   280095777 &                 HD~19782 &              A1IV/V &       9.5 &$     6.46(2)^1$ &   1.7862(1) &                                                   &                                         \\
   281668790 &                  HD~3980 &            A3VpSrCr &       5.7 &$    16.09(1)^2$ &  3.95165(9) &          3.9517(1)$^{\rm b}$, 27.151800$^{\rm g}$ &                            DD$^{\rm q}$ \\
   284196481 &                 HD~54558 &                 A1V &       7.8 &$    0.113(9)^1$ &  0.33273(4) &                                                   &                                         \\
   287329624 &                 HD~57642 &              A8IV/V &       8.5 &$     0.31(1)^1$ &    7.529(8) &                                                   &                                         \\
   287428184 &                 HD~69784 &                 A0V &       8.7 &$     4.76(2)^2$ &  0.65973(1) &                                                   &                                         \\
   289731700 &                 HD~15144 &          A3VpSrCrEu &       5.9 &$     2.43(2)^1$ &    2.992(1) &                                   2.998$^{\rm e}$ &                            DD$^{\rm r}$ \\
   300162713 &         TYC 9179-01409-1 &                   ~ &       9.2 &$     0.76(3)^1$ &    1.623(2) &                                                   &                                         \\
   301345974 &                 HD~21799 &            ApSrCrEu &       9.3 &$     13.7(1)^1$ &    5.069(3) &                                   5.121$^{\rm e}$ &                                         \\
   301481939 &                 HD~22378 &                ApSi &       9.3 &$     5.24(3)^1$ &   2.2051(5) &                                                   &                                         \\
   301795354 &                HD~204367 &          A(pSrEuCr) &       7.8 &$     0.14(1)^1$ &     9.17(8) &                                                   &                                         \\
   304096024 &                 HD~11346 &            ApSrEuCr &       9.9 &$     3.16(3)^2$ &    7.049(2) &                                7.116(6)$^{\rm b}$ &                                         \\
   304101379 &                 HD~11620 &               A4III &       8.6 &$     8.36(4)^2$ &   3.2123(4) &                                                   &                                         \\
   306573201 &                 HD~66195 &            ApSrCrEu &       8.7 &$     6.74(3)^1$ &   4.8982(6) &                             4.88938(63)$^{\rm b}$ &                                         \\
   306893839 &                 HD~68561 &                ApSi &       8.0 &$    17.40(2)^1$ &   4.2340(2) &            4.233$^{\rm e}$, 4.23415(16)$^{\rm b}$ &                                         \\
   307288162 &                 HD~71006 &                ApSi &       9.2 &$     7.00(3)^1$ &   1.5207(1) &                             1.52073(26)$^{\rm b}$ &                                         \\
   307642246 &                 HD~72634 &            ApCrEuSr &       7.3 &$     6.63(5)^2$ &   1.8608(3) &              1.8607(2)$^{\rm b}$, 0.931$^{\rm e}$ &                                    AMod \\
   307784098 &                 HD~73373 &                 A0V &       8.2 &$     0.11(1)^1$ &     5.17(3) &                                                   &                                         \\
   308085294 &                 HD~74388 &                ApSi &       7.0 &$     1.31(1)^1$ &    4.318(3) &                              4.3063(19)$^{\rm b}$ &                                    AMod \\
   309148260 &                 HD~69862 &            ApSrEuCr &      10.1 &$    13.85(3)^1$ &   13.291(7) &           13.3519(107)$^{\rm b}$, 0.519$^{\rm e}$ &                                         \\
   309402106 &                 HD~70623 &                A8IV &       8.7 &$     1.99(3)^1$ &    5.600(5) &                                                   &                          $\delta$~Scuti \\
   309792043 &                 HD~35402 &                 A3V &       7.1 &$    0.558(9)^1$ &    4.096(5) &                               15.491900$^{\rm g}$ &                                         \\
   316913639 &                HD~222638 &            ApSrEuCr &       8.7 &$     3.35(2)^2$ &   2.3469(1) &                             2.34691(26)$^{\rm b}$ &                                         \\
   326185137 &                  HD~6532 &                A2Vp &       8.4 &$     6.79(2)^1$ &   1.9451(2) &                                   1.945$^{\rm e}$ &             roAp$^{\rm s}$ MD$^{\rm d}$ \\
   326358579 &                HD~206497 &                 A3V &       8.6 &$     0.15(1)^2$ &    1.472(2) &                                                   &                                         \\
   327597288 &                HD~206653 &                ApSi &       7.2 &$    29.76(4)^1$ &  1.78698(6) &           1.787$^{\rm e}$, 1.786898(58)$^{\rm b}$ &                            ND$^{\rm d}$ \\
   327724630 &                HD~209468 &                 A1V &       7.6 &$     0.06(1)^1$ &    1.189(3) &                                                   &                                         \\
   332518087 &                HD~220455 &                 A0V &       7.8 &$     0.62(2)^1$ &    3.705(8) &                                                   &                                    AMod \\
   336731635 &                HD~214985 &                ApSi &      11.1 &$     1.25(5)^2$ &    2.770(3) &           2.77342(219)$^{\rm b}$, 1.385$^{\rm e}$ &                                         \\
   339673256 &                 HD~58292 &                ApSi &       7.9 &$    16.04(1)^1$ &   2.9607(2) &                                   2.960$^{\rm e}$ &                                         \\
   340006157 &                 HD~60435 &            ApSr(Eu) &       8.9 &$     5.66(9)^1$ &    3.805(6) &                               7.6793(6)$^{\rm t}$ &             roAp$^{\rm l}$ DD$^{\rm d}$ \\
   348898673 &                 HD~54399 &          ApSr(CrEu) &       9.7 &$     6.47(3)^1$ &   5.0045(5) &             4.9910(11)$^{\rm b}$, 2.501$^{\rm e}$ &                                         \\
   349409844 &                 HD~58448 &                ApSi &       6.9 &$    5.062(8)^2$ &  0.83090(1) &          0.831(3)$^{\rm u}$, 0.83088(5)$^{\rm b}$ &                            ND$^{\rm d}$ \\
   349972600 &                 HD~61968 &                 A3V &       9.2 &$     0.38(2)^1$ &  0.24544(2) &                                                   &                          $\delta$~Scuti \\
   350146296 &                 HD~63087 &                A7IV &       9.4 &$     1.31(2)^1$ &   2.6637(2) &                                                   &                          roAp$^{\rm b}$ \\
   350146577 &                 HD~63204 &                ApSi &       8.3 &$    48.57(1)^1$ & 1.837488(9) &                                   1.838$^{\rm e}$ &                                         \\
   350519062 &                 HD~38719 &            ApCrSrEu &       7.5 &$     6.07(2)^1$ &   4.0232(4) &           4.0237(4)$^{\rm b}$, 4.021070$^{\rm g}$ &                                         \\
   358467700 &                 HD~65712 &                ApSi &       9.3 &$    14.11(4)^1$ &   1.9457(1) &            1.946$^{\rm e}$, 1.94639(54)$^{\rm b}$ &                            DD$^{\rm d}$ \\
   364323133 &                 HD~39979 &                A6IV &       7.9 &$     1.29(3)^1$ &    4.103(7) &                                                   &                          $\delta$~Scuti \\
   364424408 &                 HD~30374 &            ApSrEuCr &      10.1 &$     8.63(3)^1$ &  1.55632(6) &            1.556$^{\rm e}$, 1.55682(14)$^{\rm b}$ &                                         \\
   365332561 &                 HD~19228 &                  A0 &       9.3 &$     1.40(3)^1$ &    2.115(2) &                                                   &                                         \\
   382044382 &                 HD~34870 &                 A2V &       9.8 &$     2.66(2)^1$ &   2.5652(5) &                                                   &                                    AMod \\
   382512330 &                 HD~64369 &                ApSi &       8.8 &$     9.01(3)^1$ &  0.89112(2) &              0.891$^{\rm e}$, 0.8912(1)$^{\rm b}$ &                                         \\
   387515681 &         TYC 0640-00521-1 &                  A5 &       9.4 &$     3.49(3)^1$ &    3.387(2) &                                                   &                                         \\
   389922504 &                 HD~40277 &       ApSrCr(Eu)pec &       8.3 &$     6.27(1)^2$ & 0.849584(6) &                             0.849585(8)$^{\rm b}$ &                                         \\
   391927730 &                 HD~56981 &                ApSr &       9.6 &$     0.44(2)^1$ &    3.788(4) &                              3.7843(18)$^{\rm b}$ &                                         \\
   392761412 &                HD~207259 &            ApEuSrCr &       8.8 &$     8.15(3)^1$ &   2.1558(2) &              2.1557(2)$^{\rm b}$, 2.200$^{\rm e}$ &                                         \\
   394230660 &                 HD~20434 &                 A3V &       9.5 &$     0.17(2)^1$ &     5.66(6) &                                                   &                          $\delta$~Scuti \\
   396696863 &                 HD~27952 &            A4mA7-A9 &       9.4 &$     2.79(3)^1$ &   0.8218(1) &                                                   &                                    AMod \\
   410451752 &                 HD~66318 &           A0pEuCrSr &       9.7 &$     0.31(4)^1$ &    0.777(1) &                             0.77688(52)$^{\rm b}$ &                            DD$^{\rm d}$ \\
   423663684 &                  HD~2957 &              ApCrEu &       8.5 &$    10.48(2)^1$ &    4.633(1) &                                   4.633$^{\rm e}$ &                            DD$^{\rm f}$ \\
   431380369 &                 HD~20880 &          ApSr(EuCr) &       8.0 &$     2.83(3)^1$ &    5.221(6) &                              5.2434(26)$^{\rm b}$ &                          roAp$^{\rm b}$ \\
   469948764 &                  HD~6208 &                 A0V &       9.4 &$     0.85(2)^1$ &   1.3006(7) &                                   4.433$^{\rm e}$ &                                         \\
\end{mpsupertabular*}

\twocolumn

\clearpage

\clearpage

\onecolumn
\tablecaption{Fundamental parameters associated with the 134 identified high probability candidate rotational variables. 
  Columns (1) through (7) list the TIC identifiers, distances ($d$), effective temperatures listed in the TIC 
  ($T_{\rm eff,TIC}$), effective temperatures ($T_{\rm eff,SED}$), luminosities ($\log{L_{\rm SED}/L_\odot}$), and 
  radii ($R_{\rm SED}$) derived through the SED fitting analysis, and the masses ($M$) derived from comparisons with 
  evolutionary models.}\label{tbl:fund_param_full}
\tablefirsthead{
    \hline
    \hline
    \noalign{\vskip0.5mm}
    TIC  & $d$  & $T_{\rm eff,TIC}$ & $T_{\rm eff,SED}$ & $\log(L_{\rm SED}/L_\odot)$ & $R_{\rm SED}$ & $M_{\rm SED}$ \\
         & (pc) & (K)               & (K)               &                             & $(R_\odot)$   & $(M_\odot)$   \\
    (1)  & (2)  & (3)               & (4)               & (5)                         & (6)           & (7)           \\
    \noalign{\vskip0.5mm}
    \hline
    \noalign{\vskip0.5mm}
}
\tablehead{
    \multicolumn{7}{l}{continued from previous page}\\
    \hline
    \hline
    \noalign{\vskip0.5mm}
    TIC  & $d$  & $T_{\rm eff,TIC}$ & $T_{\rm eff,SED}$ & $\log(L_{\rm SED}/L_\odot)$ & $R_{\rm SED}$ & $M_{\rm SED}$ \\
         & (pc) & (K)               & (K)               &                             & $(R_\odot)$   & $(M_\odot)$   \\
    (1)  & (2)  & (3)               & (4)               & (5)                         & (6)           & (7)           \\
    \noalign{\vskip0.5mm}
    \hline  
    \noalign{\vskip0.5mm}
}
\tabletail{
    \noalign{\vskip0.5mm}\hline
    \multicolumn{7}{l}{continued on next page}\\
}
\tablelasttail{
    \noalign{\vskip0.5mm}\hline
}
\centering
\begin{mpsupertabular*}{1.0\columnwidth}{@{\extracolsep{\fill}}l c c c c c r@{\extracolsep{\fill}}}
     7624182 &      $461\pm7$ &   $8666\pm229$ &   $8450\pm200$ &  $1.68\pm0.04$ &  $3.24\pm0.15$ &  $2.42\pm0.15$ \\
     7780491 &      $218\pm2$ &   $8030\pm220$ &   $7700\pm100$ &  $1.29\pm0.01$ &  $2.48\pm0.05$ &  $1.96\pm0.12$ \\
    10863314 &      $256\pm5$ &   $8968\pm234$ &   $9050\pm100$ &  $1.69\pm0.02$ &  $2.85\pm0.08$ &  $2.45\pm0.14$ \\
    12359289 &     $672\pm31$ &                &  $12300\pm250$ &  $2.56\pm0.07$ &  $4.19\pm0.28$ &  $3.95\pm0.20$ \\
    12393823 &      $226\pm5$ &   $9433\pm241$ &  $10250\pm100$ &  $1.45\pm0.03$ &  $1.68\pm0.05$ &  $2.27\pm0.09$ \\
    24186142 &      $272\pm6$ &                &  $11200\pm150$ &  $1.95\pm0.04$ &  $2.51\pm0.08$ &  $2.94\pm0.18$ \\
    24225890 &      $390\pm8$ &   $7163\pm207$ &   $7300\pm100$ &  $1.08\pm0.02$ &  $2.18\pm0.06$ &  $1.75\pm0.12$ \\
    27985664 &     $872\pm75$ &                &  $11550\pm200$ &  $2.07\pm0.10$ &  $2.69\pm0.27$ &  $3.12\pm0.20$ \\
    29432990 &      $224\pm4$ &   $7115\pm207$ &   $6750\pm100$ &  $0.89\pm0.02$ &  $2.03\pm0.07$ &  $1.58\pm0.11$ \\
    29666185 &      $123\pm1$ &   $7975\pm219$ &   $8200\pm100$ &  $1.27\pm0.02$ &  $2.14\pm0.03$ &  $1.97\pm0.13$ \\
    29755072 &      $232\pm4$ &   $8292\pm224$ &   $7800\pm100$ &  $1.54\pm0.03$ &  $3.21\pm0.09$ &  $2.23\pm0.16$ \\
    29781099 &      $204\pm1$ &   $8362\pm225$ &   $8100\pm100$ &  $1.28\pm0.02$ &  $2.22\pm0.05$ &  $1.97\pm0.12$ \\
    32035258 &      $160\pm2$ &                &  $13650\pm200$ &  $2.20\pm0.03$ &  $2.26\pm0.06$ &  $3.57\pm0.14$ \\
    38586082 &      $126\pm7$ &   $8669\pm229$ &   $9050\pm100$ &  $1.63\pm0.05$ &  $2.65\pm0.19$ &  $2.38\pm0.14$ \\
    41259805 &      $243\pm2$ &   $8034\pm220$ &   $8050\pm100$ &  $1.07\pm0.01$ &  $1.76\pm0.03$ &  $1.79\pm0.11$ \\
    42055368 &      $174\pm1$ &   $7362\pm210$ &   $7600\pm100$ &  $1.11\pm0.01$ &  $2.07\pm0.06$ &  $1.79\pm0.11$ \\
    44627561 &     $408\pm11$ &   $6451\pm198$ &   $6700\pm100$ &  $1.42\pm0.03$ &  $3.79\pm0.18$ &  $1.94\pm0.15$ \\
    44678216 &   $93.6\pm1.9$ &   $12503\pm87$ &  $13000\pm350$ &  $2.38\pm0.04$ &  $3.06\pm0.15$ &  $3.70\pm0.16$ \\
    44889961 &      $442\pm8$ &   $8823\pm232$ &   $8150\pm100$ &  $1.23\pm0.02$ &  $2.07\pm0.07$ &  $1.93\pm0.12$ \\
    52368859 &     $787\pm26$ &   $9076\pm235$ &   $9300\pm100$ &  $1.89\pm0.04$ &  $3.39\pm0.14$ &  $2.72\pm0.16$ \\
    55400261 &      $183\pm1$ &   $7173\pm208$ &   $7000\pm100$ &  $0.90\pm0.01$ &  $1.91\pm0.03$ &  $1.60\pm0.10$ \\
    66646031 &      $250\pm5$ &   $9100\pm236$ &   $8900\pm100$ &  $1.65\pm0.03$ &  $2.81\pm0.09$ &  $2.40\pm0.14$ \\
    67650835 &      $241\pm3$ &   $8944\pm233$ &   $8650\pm100$ &  $1.30\pm0.02$ &  $1.98\pm0.07$ &  $2.02\pm0.13$ \\
    70525154 &   $96.8\pm1.7$ &    $9977\pm92$ &  $10000\pm100$ &  $1.90\pm0.02$ &  $2.96\pm0.08$ &  $2.76\pm0.16$ \\
    79272047 &      $170\pm1$ &    $6782\pm70$ &   $6850\pm100$ &  $0.83\pm0.01$ &  $1.84\pm0.05$ &  $1.54\pm0.10$ \\
    79394646 &   $29.4\pm0.2$ &   $8655\pm229$ &   $8350\pm250$ &  $1.06\pm0.02$ &  $1.61\pm0.13$ &  $1.82\pm0.10$ \\
    89545031 &      $101\pm3$ &                &  $13150\pm250$ &  $2.17\pm0.06$ &  $2.35\pm0.11$ &  $3.48\pm0.11$ \\
    92705248 &     $273\pm10$ &   $8808\pm231$ &   $8600\pm100$ &  $1.15\pm0.04$ &  $1.68\pm0.08$ &  $1.90\pm0.11$ \\
   102090493 &      $360\pm4$ &   $6830\pm203$ &   $7100\pm150$ &  $1.24\pm0.03$ &  $2.76\pm0.08$ &  $1.90\pm0.13$ \\
   129636548 &     $120\pm10$ &                &  $11450\pm300$ &  $2.01\pm0.09$ &  $2.57\pm0.31$ &  $3.04\pm0.20$ \\
   136843852 &      $137\pm1$ &   $7626\pm214$ &   $7600\pm100$ &  $1.16\pm0.02$ &  $2.20\pm0.03$ &  $1.84\pm0.11$ \\
   140044682 &      $276\pm5$ &   $9726\pm246$ &   $9700\pm100$ &  $1.36\pm0.03$ &  $1.69\pm0.05$ &  $2.18\pm0.09$ \\
   140204398 &      $250\pm2$ &   $9400\pm241$ &  $10000\pm200$ &  $2.07\pm0.02$ &  $3.63\pm0.15$ &  $3.02\pm0.18$ \\
   141028198 &     $782\pm21$ &   $8192\pm222$ &   $7700\pm100$ &  $1.72\pm0.04$ &  $4.09\pm0.17$ &  $2.30\pm0.19$ \\
   141610473 &     $423\pm12$ &   $7183\pm208$ &   $6900\pm100$ &  $1.25\pm0.03$ &  $2.94\pm0.12$ &  $1.90\pm0.15$ \\
   144069014 &      $124\pm1$ &   $7520\pm212$ &   $7600\pm150$ &  $1.00\pm0.03$ &  $1.82\pm0.05$ &  $1.71\pm0.11$ \\
   147086189 &      $204\pm2$ &   $7050\pm206$ &   $7100\pm100$ &  $0.82\pm0.02$ &  $1.71\pm0.04$ &  $1.55\pm0.10$ \\
   150250959 &      $226\pm3$ &   $8072\pm220$ &   $7700\pm100$ &  $1.10\pm0.03$ &  $1.99\pm0.06$ &  $1.79\pm0.11$ \\
   153742460 &      $362\pm7$ &                &  $13050\pm250$ &  $2.32\pm0.04$ &  $2.83\pm0.09$ &  $3.63\pm0.15$ \\
   155945483 &      $197\pm2$ &   $7909\pm218$ &   $7950\pm150$ &  $1.26\pm0.03$ &  $2.24\pm0.06$ &  $1.94\pm0.12$ \\
   159834975 &   $54.7\pm1.7$ &   $8790\pm231$ &   $9750\pm150$ &  $1.57\pm0.04$ &  $2.14\pm0.11$ &  $2.37\pm0.16$ \\
   161270578 &   $41.4\pm0.5$ &   $7900\pm210$ &   $8550\pm250$ &  $1.73\pm0.02$ &  $3.34\pm0.28$ &  $2.48\pm0.15$ \\
   161334416 &     $431\pm18$ &   $8126\pm221$ &   $8100\pm100$ &  $1.59\pm0.04$ &  $3.17\pm0.16$ &  $2.30\pm0.16$ \\
   161570223 &      $378\pm8$ &   $8345\pm224$ &   $8350\pm150$ &  $1.29\pm0.04$ &  $2.10\pm0.08$ &  $1.99\pm0.13$ \\
   182909257 &      $285\pm5$ &                &  $12350\pm250$ &  $1.95\pm0.04$ &  $2.06\pm0.08$ &  $3.08\pm0.15$ \\
   183802606 &     $680\pm18$ &   $8754\pm231$ &   $9250\pm100$ &  $1.82\pm0.03$ &  $3.16\pm0.14$ &  $2.62\pm0.16$ \\
   183802904 &      $318\pm5$ &   $8648\pm229$ &   $8300\pm100$ &  $1.77\pm0.02$ &  $3.72\pm0.17$ &  $2.53\pm0.18$ \\
   200441087 &      $451\pm7$ &   $8683\pm229$ &   $8500\pm100$ &  $1.33\pm0.02$ &  $2.12\pm0.05$ &  $2.04\pm0.12$ \\
   200783972 &      $122\pm1$ &   $9068\pm235$ &  $10000\pm100$ &  $1.62\pm0.01$ &  $2.15\pm0.02$ &  $2.45\pm0.15$ \\
   201923258 &     $628\pm21$ &         $8936$ &   $9200\pm100$ &  $2.00\pm0.03$ &  $3.94\pm0.15$ &  $2.87\pm0.21$ \\
   204293493 &      $218\pm3$ &    $6490\pm83$ &   $6950\pm100$ &  $0.76\pm0.03$ &  $1.66\pm0.05$ &  $1.40\pm0.04$ \\
   204314449 &      $101\pm1$ &   $7880\pm218$ &                &                &                &  $1.88\pm0.22$ \\
   206648435 &     $400\pm10$ &   $8184\pm222$ &   $8350\pm100$ &  $1.24\pm0.03$ &  $1.99\pm0.07$ &  $1.95\pm0.12$ \\
   206663039 &   $62.3\pm1.8$ &                &  $10500\pm250$ &  $1.88\pm0.05$ &  $2.64\pm0.16$ &  $2.78\pm0.17$ \\
   207143419 &      $225\pm2$ &   $8123\pm221$ &   $8150\pm100$ &  $1.07\pm0.02$ &  $1.73\pm0.03$ &  $1.81\pm0.10$ \\
   207208753 &     $544\pm13$ &   $9310\pm239$ &   $9450\pm200$ &  $1.49\pm0.05$ &  $2.08\pm0.09$ &  $2.27\pm0.15$ \\
   211404370 &     $201\pm57$ &   $7544\pm213$ &   $7350\pm100$ &  $0.97\pm0.25$ &  $1.89\pm0.55$ &  $1.67\pm0.20$ \\
   219118940 &      $219\pm3$ &   $7172\pm208$ &   $7150\pm100$ &  $0.78\pm0.02$ &  $1.59\pm0.04$ &  $1.53\pm0.09$ \\
   219234021 &     $536\pm12$ &   $9531\pm243$ &  $10000\pm100$ &  $1.67\pm0.03$ &  $2.29\pm0.08$ &  $2.50\pm0.16$ \\
   219990936 &      $152\pm1$ &   $7313\pm209$ &   $7300\pm100$ &  $0.86\pm0.01$ &  $1.68\pm0.03$ &  $1.59\pm0.09$ \\
   220035931 &      $261\pm3$ &                &  $12950\pm250$ &  $2.29\pm0.05$ &  $2.77\pm0.06$ &  $3.58\pm0.14$ \\
   220414891 &      $230\pm2$ &   $8377\pm225$ &   $8300\pm150$ &  $1.10\pm0.02$ &  $1.72\pm0.04$ &  $1.84\pm0.10$ \\
   220565429 &      $209\pm1$ &   $7194\pm208$ &   $7050\pm100$ &  $1.01\pm0.01$ &  $2.13\pm0.04$ &  $1.68\pm0.11$ \\
   220570020 &      $184\pm1$ &   $6824\pm203$ &   $7100\pm100$ &  $0.67\pm0.02$ &  $1.43\pm0.03$ &  $1.48\pm0.07$ \\
   231813751 &      $377\pm8$ &                &  $14150\pm100$ &  $2.47\pm0.03$ &  $2.86\pm0.07$ &  $3.99\pm0.13$ \\
   231844926 &      $194\pm2$ &                &  $12700\pm350$ &  $2.11\pm0.05$ &  $2.34\pm0.07$ &  $3.33\pm0.14$ \\
   231864288 &      $140\pm1$ &   $9232\pm238$ &   $9400\pm150$ &  $1.34\pm0.03$ &  $1.76\pm0.03$ &  $2.13\pm0.11$ \\
   232066526 &     $721\pm22$ &   $8750\pm230$ &   $8700\pm300$ &  $1.31\pm0.05$ &  $1.98\pm0.14$ &  $2.04\pm0.13$ \\
   234346165 &     $726\pm24$ &         $9480$ &  $10700\pm100$ &  $2.20\pm0.04$ &  $3.68\pm0.15$ &  $3.25\pm0.15$ \\
   235007556 &      $124\pm2$ &                &  $14750\pm250$ &  $2.31\pm0.04$ &  $2.18\pm0.05$ &  $3.84\pm0.25$ \\
   237336864 &      $245\pm2$ &   $8283\pm223$ &   $8100\pm100$ &  $0.91\pm0.02$ &  $1.45\pm0.04$ &  $1.62\pm0.06$ \\
   259587315 &      $355\pm5$ &   $8165\pm222$ &   $7750\pm100$ &  $1.41\pm0.02$ &  $2.83\pm0.11$ &  $2.09\pm0.13$ \\
   262613883 &      $402\pm6$ &   $9359\pm240$ &   $9100\pm150$ &  $1.44\pm0.03$ &  $2.11\pm0.06$ &  $2.19\pm0.14$ \\
   269857621 &      $196\pm2$ &   $9136\pm236$ &   $8700\pm100$ &  $1.06\pm0.02$ &  $1.50\pm0.03$ &  $1.72\pm0.07$ \\
   270250508 &      $151\pm1$ &    $6973\pm76$ &   $6850\pm100$ &  $0.92\pm0.01$ &  $2.05\pm0.05$ &  $1.62\pm0.11$ \\
   270304671 &      $387\pm6$ &   $8044\pm220$ &   $7750\pm100$ &  $1.22\pm0.02$ &  $2.27\pm0.08$ &  $1.90\pm0.11$ \\
   270406421 &      $117\pm1$ &   $7918\pm218$ &   $7700\pm100$ &  $1.37\pm0.01$ &  $2.73\pm0.05$ &  $2.05\pm0.13$ \\
   272316843 &      $319\pm4$ &   $8547\pm227$ &   $8150\pm150$ &  $1.17\pm0.03$ &  $1.93\pm0.06$ &  $1.88\pm0.12$ \\
   277688819 &      $115\pm2$ &   $8368\pm225$ &   $8200\pm100$ &  $1.13\pm0.02$ &  $1.83\pm0.08$ &  $1.85\pm0.11$ \\
   277748932 &     $518\pm14$ &   $8955\pm234$ &   $8500\pm100$ &  $1.33\pm0.03$ &  $2.12\pm0.08$ &  $2.04\pm0.13$ \\
   278804454 &      $146\pm3$ &   $8672\pm229$ &   $8450\pm100$ &  $1.48\pm0.02$ &  $2.56\pm0.08$ &  $2.18\pm0.13$ \\
   279573219 &   $85.4\pm1.4$ &                &  $11250\pm250$ &  $1.93\pm0.04$ &  $2.42\pm0.10$ &  $2.92\pm0.18$ \\
   280095777 &      $327\pm5$ &   $8808\pm231$ &   $8750\pm150$ &  $1.13\pm0.03$ &  $1.60\pm0.06$ &  $1.91\pm0.09$ \\
   281668790 &   $68.3\pm0.4$ &   $8747\pm230$ &   $8500\pm150$ &  $1.28\pm0.02$ &  $2.01\pm0.08$ &  $2.00\pm0.13$ \\
   284196481 &      $257\pm2$ &   $8650\pm229$ &   $8650\pm100$ &  $1.62\pm0.02$ &  $2.86\pm0.05$ &  $2.35\pm0.13$ \\
   287329624 &      $135\pm1$ &    $6900\pm94$ &   $7000\pm100$ &  $0.76\pm0.01$ &  $1.63\pm0.02$ &  $1.50\pm0.10$ \\
   287428184 &      $261\pm3$ &   $9259\pm238$ &   $9000\pm100$ &  $1.30\pm0.02$ &  $1.83\pm0.03$ &  $2.05\pm0.12$ \\
   289731700 &   $81.7\pm1.0$ &   $8429\pm226$ &   $8450\pm100$ &  $1.37\pm0.02$ &  $2.27\pm0.06$ &  $2.08\pm0.13$ \\
   300162713 &      $288\pm4$ &   $8826\pm232$ &   $8500\pm100$ &  $1.14\pm0.02$ &  $1.72\pm0.05$ &  $1.89\pm0.10$ \\
   301345974 &      $305\pm4$ &   $8212\pm222$ &   $7400\pm100$ &  $1.12\pm0.02$ &  $2.21\pm0.09$ &  $1.79\pm0.11$ \\
   301481939 &     $482\pm13$ &   $9501\pm242$ &  $11150\pm200$ &  $1.77\pm0.05$ &  $2.05\pm0.07$ &  $2.73\pm0.16$ \\
   301795354 &      $168\pm2$ &   $8790\pm231$ &   $8450\pm100$ &  $1.23\pm0.02$ &  $1.93\pm0.04$ &  $1.95\pm0.13$ \\
   304096024 &     $751\pm20$ &   $8098\pm221$ &   $7600\pm100$ &  $1.65\pm0.03$ &  $3.88\pm0.14$ &  $2.21\pm0.19$ \\
   304101379 &      $233\pm2$ &   $8046\pm220$ &   $7650\pm100$ &  $1.19\pm0.01$ &  $2.23\pm0.05$ &  $1.86\pm0.12$ \\
   306573201 &      $292\pm8$ &                &   $8950\pm100$ &  $1.42\pm0.04$ &  $2.14\pm0.08$ &  $2.16\pm0.14$ \\
   306893839 &      $516\pm9$ &                &  $12700\pm200$ &  $2.47\pm0.04$ &  $3.55\pm0.12$ &  $3.82\pm0.16$ \\
   307288162 &     $809\pm30$ &                &  $13600\pm200$ &  $2.42\pm0.05$ &  $2.92\pm0.16$ &  $3.83\pm0.15$ \\
   307642246 &     $353\pm36$ &   $8947\pm233$ &   $9650\pm100$ &  $2.19\pm0.09$ &  $4.48\pm0.51$ &  $2.99\pm0.21$ \\
   307784098 &      $278\pm3$ &   $8396\pm225$ &   $8550\pm100$ &  $1.56\pm0.02$ &  $2.75\pm0.05$ &  $2.28\pm0.13$ \\
   308085294 &      $349\pm5$ &                &  $13850\pm200$ &  $2.62\pm0.04$ &  $3.54\pm0.16$ &  $4.25\pm0.17$ \\
   309148260 &     $632\pm11$ &   $8134\pm221$ &   $7350\pm100$ &  $1.43\pm0.03$ &  $3.20\pm0.09$ &  $2.10\pm0.17$ \\
   309402106 &      $158\pm1$ &   $7053\pm206$ &   $7050\pm100$ &  $0.76\pm0.01$ &  $1.60\pm0.02$ &  $1.51\pm0.09$ \\
   309792043 &      $164\pm2$ &   $8482\pm226$ &   $8450\pm100$ &  $1.52\pm0.03$ &  $2.69\pm0.06$ &  $2.23\pm0.13$ \\
   316913639 &      $307\pm5$ &   $9714\pm245$ &   $9550\pm250$ &  $1.47\pm0.03$ &  $1.98\pm0.09$ &  $2.25\pm0.13$ \\
   326185137 &      $197\pm3$ &   $8550\pm227$ &   $8450\pm100$ &  $1.11\pm0.02$ &  $1.67\pm0.05$ &  $1.86\pm0.09$ \\
   326358579 &      $215\pm3$ &   $7573\pm213$ &   $7500\pm100$ &  $1.12\pm0.03$ &  $2.15\pm0.05$ &  $1.79\pm0.12$ \\
   327597288 &      $240\pm5$ &                &  $12100\pm500$ &  $2.08\pm0.07$ &  $2.49\pm0.15$ &  $3.20\pm0.18$ \\
   327724630 &      $162\pm2$ &   $8906\pm233$ &   $9350\pm100$ &  $1.39\pm0.01$ &  $1.88\pm0.03$ &  $2.16\pm0.11$ \\
   332518087 &      $201\pm2$ &   $9631\pm244$ &  $10100\pm250$ &  $1.52\pm0.03$ &  $1.87\pm0.07$ &  $2.36\pm0.12$ \\
   336731635 &   $1640\pm270$ &   $9439\pm241$ &  $12150\pm550$ &  $2.28\pm0.19$ &  $3.13\pm0.67$ &  $3.50\pm0.17$ \\
   339673256 &      $409\pm6$ &   $9601\pm244$ &  $10450\pm250$ &  $2.12\pm0.04$ &  $3.52\pm0.15$ &  $3.12\pm0.18$ \\
   340006157 &      $245\pm2$ &   $8427\pm226$ &   $7800\pm100$ &  $1.10\pm0.01$ &  $1.93\pm0.08$ &  $1.79\pm0.11$ \\
   348898673 &     $700\pm16$ &    $7505\pm44$ &   $7300\pm100$ &  $1.67\pm0.03$ &  $4.26\pm0.17$ &  $2.12\pm0.04$ \\
   349409844 &      $159\pm1$ &                &  $11500\pm200$ &  $1.78\pm0.03$ &  $1.97\pm0.04$ &  $2.79\pm0.13$ \\
   349972600 &      $281\pm3$ &   $9540\pm243$ &   $7800\pm200$ &  $1.08\pm0.03$ &  $1.91\pm0.06$ &  $1.78\pm0.11$ \\
   350146296 &      $221\pm3$ &   $7690\pm215$ &   $7200\pm100$ &  $0.79\pm0.02$ &  $1.59\pm0.04$ &  $1.54\pm0.09$ \\
   350146577 &      $336\pm5$ &   $9737\pm246$ &  $10700\pm150$ &  $1.83\pm0.03$ &  $2.38\pm0.06$ &  $2.74\pm0.17$ \\
   350519062 &      $223\pm2$ &   $8937\pm233$ &   $8950\pm100$ &  $1.62\pm0.02$ &  $2.70\pm0.05$ &  $2.37\pm0.14$ \\
   358467700 &      $432\pm6$ &   $8768\pm231$ &  $10000\pm350$ &  $1.56\pm0.04$ &  $2.01\pm0.11$ &  $2.39\pm0.14$ \\
   364323133 &      $118\pm1$ &   $7578\pm213$ &   $7250\pm100$ &  $0.86\pm0.02$ &  $1.70\pm0.04$ &  $1.58\pm0.09$ \\
   364424408 &     $643\pm21$ &    $7608\pm47$ &   $8100\pm150$ &  $1.49\pm0.05$ &  $2.83\pm0.14$ &  $1.95\pm0.18$ \\
   365332561 &      $311\pm8$ &   $9470\pm242$ &   $8500\pm100$ &  $1.17\pm0.03$ &  $1.78\pm0.06$ &  $1.91\pm0.12$ \\
   382044382 &      $370\pm6$ &   $7917\pm218$ &   $7950\pm150$ &  $1.09\pm0.03$ &  $1.86\pm0.07$ &  $1.80\pm0.12$ \\
   382512330 &      $430\pm6$ &   $9042\pm235$ &   $9350\pm100$ &  $1.72\pm0.02$ &  $2.76\pm0.05$ &  $2.50\pm0.15$ \\
   387515681 &                &     $6885\pm9$ &   $6600\pm100$ &                &                &                \\
   389922504 &      $207\pm1$ &                &   $8750\pm100$ &  $1.20\pm0.01$ &  $1.73\pm0.03$ &  $1.95\pm0.09$ \\
   391927730 &      $230\pm1$ &   $6984\pm205$ &   $6750\pm100$ &  $0.76\pm0.01$ &  $1.75\pm0.02$ &  $1.48\pm0.11$ \\
   392761412 &      $304\pm5$ &   $8036\pm220$ &   $7850\pm100$ &  $1.31\pm0.02$ &  $2.45\pm0.07$ &  $1.98\pm0.11$ \\
   394230660 &      $320\pm4$ &   $7908\pm218$ &   $7950\pm200$ &  $1.10\pm0.03$ &  $1.86\pm0.06$ &  $1.81\pm0.11$ \\
   396696863 &      $208\pm1$ &   $6562\pm199$ &   $6800\pm100$ &  $0.80\pm0.02$ &  $1.81\pm0.03$ &  $1.52\pm0.11$ \\
   410451752 &     $509\pm14$ &   $9057\pm235$ &   $8900\pm100$ &  $1.52\pm0.03$ &  $2.41\pm0.10$ &  $2.25\pm0.14$ \\
   423663684 &      $327\pm9$ &         $9185$ &   $9400\pm100$ &  $1.58\pm0.03$ &  $2.31\pm0.10$ &  $2.35\pm0.15$ \\
   431380369 &      $221\pm5$ &   $8242\pm223$ &   $8000\pm100$ &  $1.38\pm0.02$ &  $2.56\pm0.12$ &  $2.06\pm0.12$ \\
   469948764 &      $327\pm7$ &   $8779\pm231$ &   $8950\pm100$ &  $1.22\pm0.03$ &  $1.69\pm0.06$ &  $1.99\pm0.11$ \\
\end{mpsupertabular*}

\twocolumn

\clearpage

\end{document}